# High efficiency and compact lithium niobate non-resonant recirculating phase modulator and its applications

Feiyu Wang[1,†], Liheng Wang[1,†], Mingrui Yuan[1,†], Zhen Han[1], Binjie Wang[1], Yong Zheng[1], Pu Zhang[1], Yongheng Jiang[1], Huifu Xiao[1], Mei Xian Low[2], Aditya Dubey[2], Thach Giang Nguyen[2], Guanghui Ren[2], Arnan Mitchell[2], and Yonghui Tian[1,3*]

F. Wang, L. Wang, M. Yuan, Z. Han, B. Wang, Y. Zheng, P. Zhang, Y. Jiang, H. Xiao, Y. Tian
1. School of Physical Science and Technology, Lanzhou University, Lanzhou 730000, Gansu, China

M. X. Low, A. Dubey, T. G. Nguyen, G. Ren, A. Mitchell
2. Integrated Photonics and Applications Centre (InPAC), School of Engineering, RMIT University, Melbourne, VIC 3001, Australia

Y. Tian
3. School of Mathematics and Physics, North China Electric Power University, Beijing 102206, China

† These authors contributed equally.
*Corresponding author: tianyh@lzu.edu.cn

**Abstract:** High modulation efficiency and a compact footprint are critical for next-generation electro-optic (EO) modulators. We introduce a new class of non-resonant recirculating phase modulators (PMs) that boosts modulation efficiency by repeatedly modulating the optical field within a single, non-resonant waveguide, while fundamentally removing the loop-length matching constraint that has limited prior recirculating schemes. This architectural breakthrough simultaneously enables a much smaller device footprint and an extended low-$V_\pi$ bandwidth, without relying on narrowband resonances. Building on this concept, we experimentally demonstrate both a Mach–Zehnder modulator (MZM) and a cascaded PM, and verify their versatility in finite impulse response (FIR) filtering and optical frequency comb (OFC) generation. The recirculating MZM operates as a 4-tap rectangular-window FIR filter with 110 GHz bandwidth in a compact 2.889×0.58 $mm^2$ footprint. The cascaded PM achieves a 3.40 GHz low-$V_\pi$ bandwidth, a 110 GHz resonant EO bandwidth, and a $V_\pi L$ of 0.7 V·cm, and generates 20 OFC lines under a 33 dBm microwave drive. These results demonstrate, for the first time, a practical and highly efficient non-resonant recirculating modulation platform, laying the groundwork for scalable high-order mode recirculating modulators (RMs) and opening new opportunities in optical communications, sensing, and microwave photonics.

## 1. Introduction

Electro optic (EO) phase modulators (PMs) are indispensable building blocks in contemporary photonic systems, including radio over fiber (RoF) architectures, beyond-5G communication networks[1,2], optical frequency comb (OFC) generation[3,4], and microwave photonics (MWP)[5,6]. As optical and wireless technologies advance toward higher sensitivity, ultrahigh spectral efficiency, and increasingly broad bandwidths, PMs are required to simultaneously provide high modulation efficiency, large low-half-wave-voltage ($V_\pi$) bandwidth, and a compact footprint. Low-$V_\pi$ PMs are essential for ultrahigh spectral efficiency narrowband orthogonal frequency division multiplexing (OFDM) and high bit rate wideband single carrier transmission in millimeter wave bands[7], while PMs with moderate $V_\pi$ enable flat OFC generation with large line spacing under strong radio frequency (RF) drive power[8]. In addition, PM-integrated Fourier domain mode-locked optoelectronic oscillators (FDML OEOs) allow broadband frequency tuning across tens of gigahertz[9]. These diverse applications collectively demand PM architectures that utilize both optical and electrical resources with high efficiency. However, most conventional on chip PMs guide the optical mode through only a single electrode gap, leading to severe underutilization of the available electrode spacing. This intrinsic structural constraint fundamentally limits further improvement in modulation efficiency and effective bandwidth utilization, particularly when device footprint scaling is restricted. Overcoming this bottleneck is therefore critical for next generation high performance photonic systems.

Thin film lithium niobate (TFLN) has recently emerged as a leading platform for high performance EO modulation, owing to its broad transparency window, large EO coefficient, and low propagation loss[10-12]. Despite these advantages, conventional TFLN PMs still suffer from an unfavorable trade off among the voltage length product ($V_\pi L$), low-$V_\pi$ operating bandwidth, and device size. Representative devices exhibit $V_\pi L$ values of approximately 7 V·cm at 5 GHz[13], and even folded waveguide designs with a 5 cm modulation length achieve an RF $V_\pi$ of only 1.43 V at 3 GHz[14]. To date, no substantial reduction in $V_\pi L$ has been achieved for single channel PMs. Recirculating or recycling PMs have been proposed to enhance modulation efficiency by guiding the optical wave through multiple electrode gaps[15,16]. Although these devices demonstrate reduced $V_\pi$ values down to 1.9–3.2 V over multiple modulation periods, their performance relies on microwave resonances. As a result, the effective low-$V_\pi$ bandwidth is limited to about 1.4 GHz per modulation period[16], which is insufficient for wideband applications. Moreover, low-$V_\pi$ operation generally requires long electrodes, increasing lateral footprint and limiting integration density. Folded layouts partially mitigate this issue[14,17,18], while non-resonant recirculating modulators (RMs) provide a potential route to shortening the physical length[19]. Nevertheless, all reported non resonant designs rely on unequal loop lengths, imposing stringent loop length matching conditions that inherently couple the extra optical path length to electrode length and modulation pass number. This architectural constraint simultaneously degrades footprint efficiency and low-$V_\pi$ bandwidth, preventing a practical, highly efficient non-resonant recirculating modulation platform.

Here, we introduce a fundamentally different class of novel non-resonant recirculating modulator (NRM) on the TFLN platform that overcomes these limitations. Our approach employs slotted multimode waveguide bends (SMWBs) and six mode converters to form a mode loop structure in which light is recycled four times within a single non-resonant waveguide as $TE_0$, $TE_1$, $TE_2$, and $TE_3$ modes. By engineering the multimode bends and mode converters, the additional loop length is fully decoupled from both electrode length and mode number, thereby eliminating the loop length matching constraint that has restricted prior recirculating schemes. This architecture enables repeated non-resonant EO modulation within a compact footprint, simultaneously enhancing modulation efficiency and extending the low-$V_\pi$ bandwidth. As a proof of concept, we fabricate 1, 2, 3, and 4 loop NRMs to experimentally validate the scalability of this approach. For 3 mode and 4 mode devices with 3 mm electrodes, the extra length is only 0.730 mm and 0.889 mm, corresponding to effective modulation length percentages of 80.44% and 77.14%, respectively. Notably, the extra length remains nearly independent of both mode count and electrode length, while the effective modulation length percentage increases markedly with electrode length. We further verify substantial enhancement of low-$V_\pi$ bandwidth in the non-resonant regime. An NRM with a single modulation region and a 2 mm electrode achieves a low-$V_\pi$ bandwidth of 4.87 GHz, compared with 1.27 GHz for a conventional RM of identical length. Similarly, a two-modulation region NRM with a 1 mm electrode reaches 3.40 GHz, exceeding the 1.05 GHz bandwidth of its conventional counterpart. Building on this platform, we demonstrate both a recirculating Mach–Zehnder modulator (MZM) and a cascaded PM to highlight the versatility of the proposed NRM in system level functions. The recirculating MZM operates as a 4-tap rectangular window finite impulse response (FIR) filter with an operating bandwidth exceeding 110 GHz in a compact 2.889 × 0.58 mm² footprint, representing, to our knowledge, the largest operating bandwidth and most compact footprint reported for an integrated FIR filter. Based on the same NRM architecture, an 8-mode cascaded PM achieves, to our knowledge, the broadest low-$V_\pi$ bandwidth of 3.40 GHz, a resonant EO bandwidth beyond 110 GHz, and the lowest reported $V_\pi L$ of 0.7 V·cm among comparable devices. Under a 33 dBm microwave drive, the cascaded PM generates 20 OFC lines, demonstrating its capability as an efficient multiwavelength source. Together, these results establish a practical and scalable non-resonant recirculating modulation platform and open new opportunities for high efficiency EO modulation in next generation photonic systems.

## 2. Results

### 2.1 Principles and designs

Figure 1a illustrates the proposed three-loop NRM. The device comprises two SMWBs, six mode converters, three single-mode circular-bend waveguides, and two straight multimode waveguides. The graph of the utilized mode converters is shown in Figure 1b. Light is injected from the left single-mode input in the $TE_0$ mode and subsequently expanded into a multimode waveguide through three cascaded tapers, which sequentially enable $TE_0/TE_1$, $TE_0/TE_1/TE_2$, and $TE_0/TE_1/TE_2/TE_3$ propagation. After the tapers, most of the optical power remains in $TE_0$ and experiences its first EO modulation under traveling-wave electrodes, whose waveguide cross section is shown in Figure 1c. The modulated $TE_0$ mode is then routed through the SMWBs whose cross section is shown in Figure 1d to form a recirculation path and returned to the multimode bus. Following two-stage narrowing back to single-mode operation and a circular bend, $TE_0$ is successively converted to $TE_1$, $TE_2$, and $TE_3$ to form three loops with their optical field intensity distributions shown in Figure 1e. Each higher-order mode propagates through the same modulation region again, is routed back by the SMWBs, and is demultiplexed to $TE_0$ before conversion to the next mode. The overview of the SMWBs which assist $TE_1$, $TE_2$, and $TE_3$ is shown in Figure 1f. After the fourth pass through the modulation region as $TE_3$, the light is finally converted and coupled out as $TE_0$ into the single-mode output waveguide.

The mode converters ($TE_0/TE_1$, $TE_0/TE_2$, and $TE_0/TE_3$) and SMWBs are designed along the crystallographic *Y* direction of the TFLN platform based on mode effective indices reported in our previous work[20,21]. This architecture enables multiple non-resonant modulation passes within a single waveguide while fully decoupling the additional loop length from both electrode length and mode order. When the optical delays of the three loops are equal, constructive enhancement of modulation efficiency occurs over specific microwave frequency ranges, without introducing narrowband resonances (see Supplementary Note S1 for details). The mechanism by which the NRM circumvents the loop length matching constraint is described in Supplementary Note S2. The width and length of each part of the NRM is elaborated to achieve identical time delay of each loop (see Supplementary Note S3 for details). The SMWB consists of a silicon substrate, a 4.7 μm buried $SiO_2$ layer, an *X*-cut 300 nm $LiNbO_3$ thin film, and three 300 nm $Si_3N_4$ waveguides. The modulation section employs 3 mm long traveling-wave gold electrodes optimized for RF–optical velocity matching and impedance matching, with a signal and ground width of 50 μm and 108 μm and an electrode gap of 9.717 μm to balance low optical loss and fabrication tolerance. All SMWBs and mode converters are designed for broadband, low-loss operation, and all tapers are 30 μm long to ensure near-adiabatic mode transitions with minimal excess path length (see Supplementary Note S4 for details). The SMWBs use a compact bending radius of 120 μm and a multimode width of 4.48 μm, incorporating two 120 nm wide air slots to support stable higher-order-mode guidance. The three single-mode circular-bend waveguides on the left side have radii of 98.4 μm, 116.57 μm, and 127.39 μm, respectively, providing identical loop delays of $\Delta T = 1/16.9$ ns. The fabricated device, shown in Figure 1g–l, occupies an on-chip footprint of $3.89 \times 0.45$ mm$^2$, with the active modulation section accounting for 77.14% of the total device length.

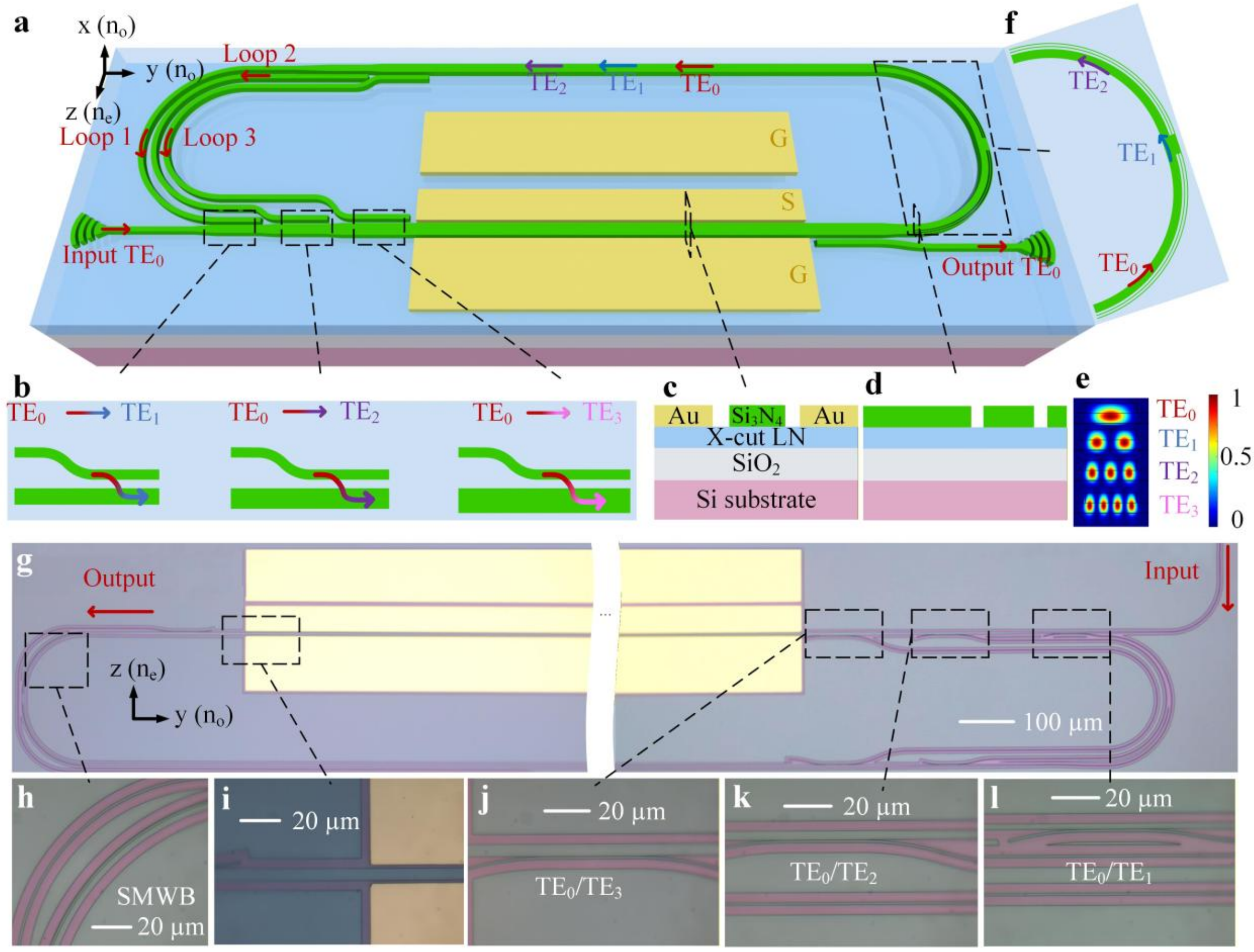

**Figure 1.** **a** Schematic overview of the proposed NRMs. **b** Structural layouts of the $TE_0/TE_1$, $TE_0/TE_2$, and $TE_0/TE_3$ mode converters. **c** Cross-sectional view of the modulation region. **d** Cross section of the SMWBs. **e** Simulated optical field intensity distributions in the multimode waveguide within the modulation region. **f** Overview of the slotted multimode waveguide bends. **g** Optical micrograph of the fabricated three-loop NRM. **h** Fabricated SMWBs supporting $TE_0$, $TE_1$, and $TE_2$ modes. **i** Fabricated traveling-wave electrodes and the input multimode waveguide. **j** Fabricated $TE_0/TE_3$ mode converter. **k** Fabricated $TE_0/TE_2$ mode converter. **l** Fabricated $TE_0/TE_1$ mode converter.

## 2.2 Performance characterization of NRM structure

To enable a systematic comparison of modulation efficiency enhancement, we experimentally evaluated four PMs using the experimental setup shown in Figure 2a, as illustrated in the inset of Figure 2a: (1) PM; (2) RM1, a single-loop RM; (3) RM2, a two-loop RM; and (4) RM3, a three-loop RM. In the first step, the RF power was carefully adjusted to equalize the modulation indices of all modulators. The exact modulation indices were extracted from the measured carrier-to-sideband power ratios. As shown in Figure 2b i–iv, all four modulators exhibit nearly identical optical spectra, despite a systematic reduction in the arbitrary waveform generator (AWG) output voltage from 500 mV to 250 mV, 166.7 mV, and 125 mV for PM, RM1, RM2, and RM3, respectively. The extracted modulation index remains nearly constant at 0.1462 ± 0.01, as summarized in Figure 2b v, in good agreement with the theoretical expectation for RF power reduction. We then evaluated modulation efficiency under a fixed RF drive condition, with an identical AWG output voltage of 700 mV applied to all modulators. The measured spectra in Figure 2c i–iv show a clear reduction in the carrier-to-sideband power difference as the number of resonant loops increases. All modulators share identical electrode geometry, including electrode length, gap, and thickness, ensuring a fair comparison. The extracted modulation indices, shown in Figure 2c v, are 0.1914 for PM, 0.3602 for RM1, 0.5019 for RM2, and 0.7659 for RM3. These results closely follow the theoretical predictions and quantitatively confirm the progressive enhancement of modulation efficiency enabled by the resonant loop architecture.

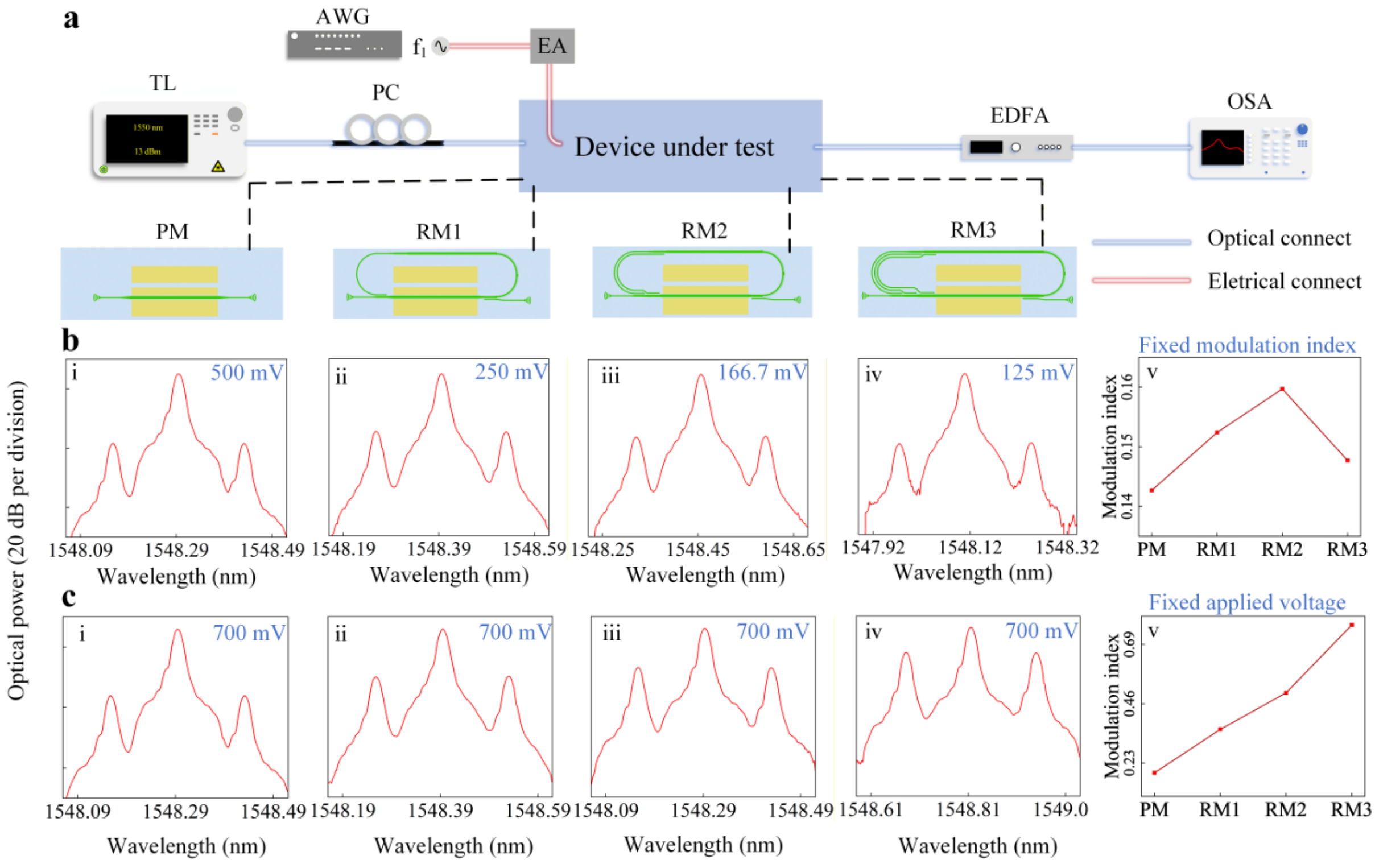

**Figure 2.** Measurement of the proposed NRM. **a** Experimental setup used for modulation characterization. **b** Measured optical spectra obtained with adjusted RF drive voltages to equalize the modulation index: i 500 mV applied to PM, ii 250 mV to RM1, iii 166.7 mV to RM2, and iv 125 mV to RM3. v The corresponding extracted modulation indices. **c** Measured optical spectra under a fixed RF drive voltage of 700 mV applied to i PM, ii RM1, iii RM2, and iv RM3. v The corresponding extracted modulation indices. TL, tunable laser; PC, polarization controller; AWG, arbitrary waveform generator; EA, electrical amplifier; EDFA, erbium doped fiber amplifier; OSA, optical spectrum analyzer; PM, phase modulator; RM1, recirculating modulator with one loop; RM2, recirculating modulator with two loops; RM3, recirculating modulator with three loops.

To analyze modulation efficiency, we measured the modulation index over a wide range of RF power levels, as shown in Figure 3a. The modulation index follows a quasi-quadratic dependence on the applied RF power, which is consistent with the voltage squared scaling of RF power. Clear efficiency enhancement is observed for the RMs. RM1, RM2, and RM3 exhibit approximately twofold, threefold, and fourfold higher modulation indices, respectively, compared with the PM that shares an identical electrode layout. As non-resonant devices, the RMs are expected to support a broad optical bandwidth, which is mainly limited by the wavelength dependent response of the mode converters and SMWBs. To quantify this bandwidth, we applied a fixed RF power that yields a constant modulation index and measured the modulation index from 1533 nm to 1563 nm with 5 nm steps. As shown in Figure 3b, the modulation index remains largely insensitive to wavelength. The weak residual variation arises from the intrinsic wavelength dependence of the mode converters and SMWBs.

The proposed NRM structure also enables a much more compact footprint compared with its conventional counterparts. To quantify this advantage, we define two metrics, namely the effective modulation length percentage and the extra length. The effective modulation length percentage is defined as the ratio of electrode length to total device length. As shown in Figure 3c, the 3 mm electrode NRM2 and NRM3 achieve effective modulation length percentages of 80.44 percent and 77.14 percent, respectively. In contrast, the 6.771 mm conventional recirculating modulator (ORM) with two loops exhibits an effective modulation length percentage of 62.44 percent, while the simulated value for a 6.771 mm NRM2 reaches 90.18 percent and can be further improved by increasing the electrode length. For ORMs, the effective modulation length percentage decreases rapidly with increasing mode count. By comparison, NRMs maintain a nearly constant and substantially higher percentage even for higher order modes. For a fixed total device length of 1 cm, the simulated effective modulation length percentage is 44.4 percent for ORM4 and 90.5 percent for NRM4, highlighting the intrinsic compactness of the NRM architecture.

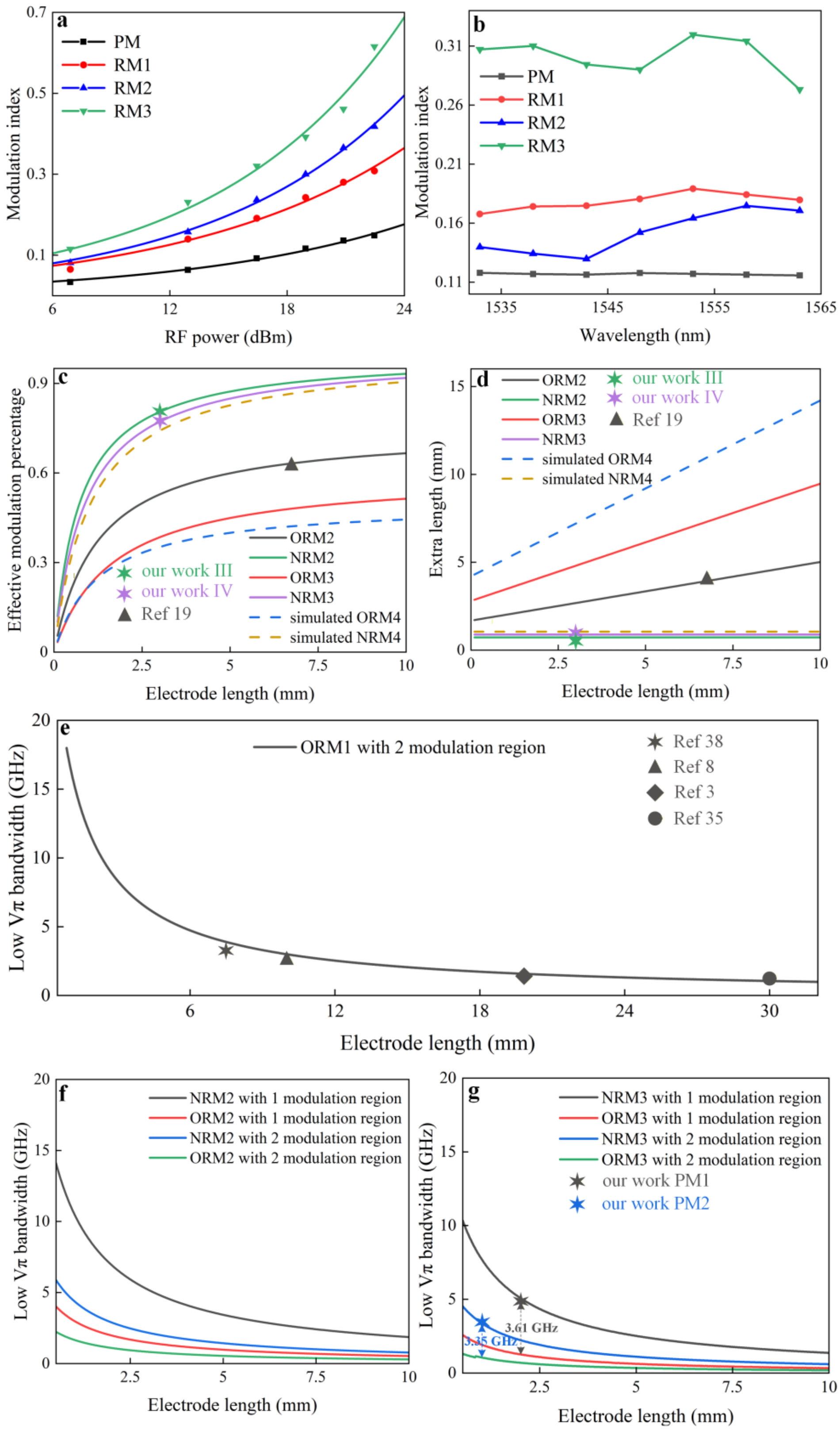

**Figure 3.** Modulation efficiency, footprint, and low-$V_\pi$ bandwidth analysis of NRMs. **a** Modulation index of PM, RM1, RM2, and RM3 as a function of the applied RF power. **b** Modulation index measured across the C band as a function of operational wavelength for different modulators. **c** Theoretical analysis of the effective modulation length percentage for the proposed and conventional RM structures. The dotted orange and blue curves indicate the simulated results for RM4, which was not fabricated. **d** Theoretical analysis of the extra length for the proposed and conventional RM structures. The dotted orange and blue curves correspond to the simulated RM4. **e**

Theoretical analysis together with representative literature data of the low-$V_\pi$ bandwidth for the RM with one loop. **f** Theoretical analysis of the low-$V_\pi$ bandwidth for the proposed and conventional RM structures with two loops. **g** Theoretical analysis and our experimental results for PM1 and PM2, illustrating the low-$V_\pi$ bandwidth of the proposed and conventional RM structures with three loops. PM, phase modulator; RM, recirculating modulator; RM1, recirculating modulator with one loop; RM2, recirculating modulator with two loops; RM3, recirculating modulator with three loops; ORM1, conventional recirculating modulator with one loop; ORM2, conventional recirculating modulator with two loops; ORM3, conventional recirculating modulator with three loops; NRM2, novel recirculating modulator with two loops; NRM3, novel recirculating modulator with three loops; PM1, 2 mm PM type I; PM2, 1 mm PM type II.

The extra length is defined as the difference between the total device length and the electrode length. Figure 3d shows that the 3 mm NRM2 and NRM3 have extra lengths of only 0.730 mm and 0.889 mm, respectively. In contrast, the 6.771 mm ORM2 exhibits an extra length of 4.079 mm, while the simulated extra length for a 6.771 mm NRM2 remains at 0.730 mm and is independent of electrode length. For ORMs, the extra length increases with both the number of modes and the electrode length. In sharp contrast, the extra length of NRMs remains nearly constant and significantly smaller, even for higher order modes or longer electrodes. At a fixed total length of 1 cm, the simulated extra lengths for ORM4 and NRM4 amount to 14.2 mm and 1.0485 mm, respectively, further underscoring the superior compactness of the NRM. To further validate the advantage of the NRM structure in enhancing the low-$V_\pi$ bandwidth, we fabricated a 2 mm PM type I (PM1) and a 1 mm PM type II (PM2). Figure 3f presents the theoretical analysis of the low-$V_\pi$ bandwidth for ORM1 with a single modulation region, together with representative results from the literature. Figure 3g shows the theoretical analysis of the low $V_\pi$ bandwidth for NRM2 and ORM2 with one and two modulation regions. Figure 3e compares these theoretical predictions with our experimental results for PM1 and PM2. Under the same electrode length, the low-$V_\pi$ bandwidth is improved by 3.63 GHz for PM1 and by 3.35 GHz for PM2, providing direct experimental confirmation of the NRM advantage.

### 2.3 The MZM based on the NRM structure

Based on the proposed structure, we designed a MZM on the TFLN platform, as shown in Figure 4a. The input light is first evenly divided into two paths by a subwavelength grating (SWG) assisted Y-branch and then propagates through a multimode waveguide. The two optical beams subsequently traverse the upper and lower NRMs, where each beam experiences four successive modulation interactions within the modulation region, as discussed above. Finally, the modulated beams are recombined by a second SWG assisted Y-branch at the output. Figure 4b presents an overview of our previously reported SWG assisted Y-branch[22]. Figure 4c shows the transverse cross section of the modulation region, in which electrodes are symmetrically placed on both sides of the multimode waveguide. Figure 4d illustrates the optical field intensity distributions of the supported modes in the multimode waveguide of the modulation region. The width of the bus waveguide is narrowed to 4.779 um, which minimizes optical loss induced by the metallic electrodes while simultaneously enhancing the modulation efficiency. The detailed structural parameters of the SWG assisted Y-branch and the modulation region are summarized in Supplementary Table 1. The fabricated MZM with six recirculation loops is shown in Figure 4e. The device occupies an on-chip area of 2.889 × 0.58 mm², achieves an effective modulation length percentage of 69.23 percent, and exhibits an extra length of 0.889 mm.

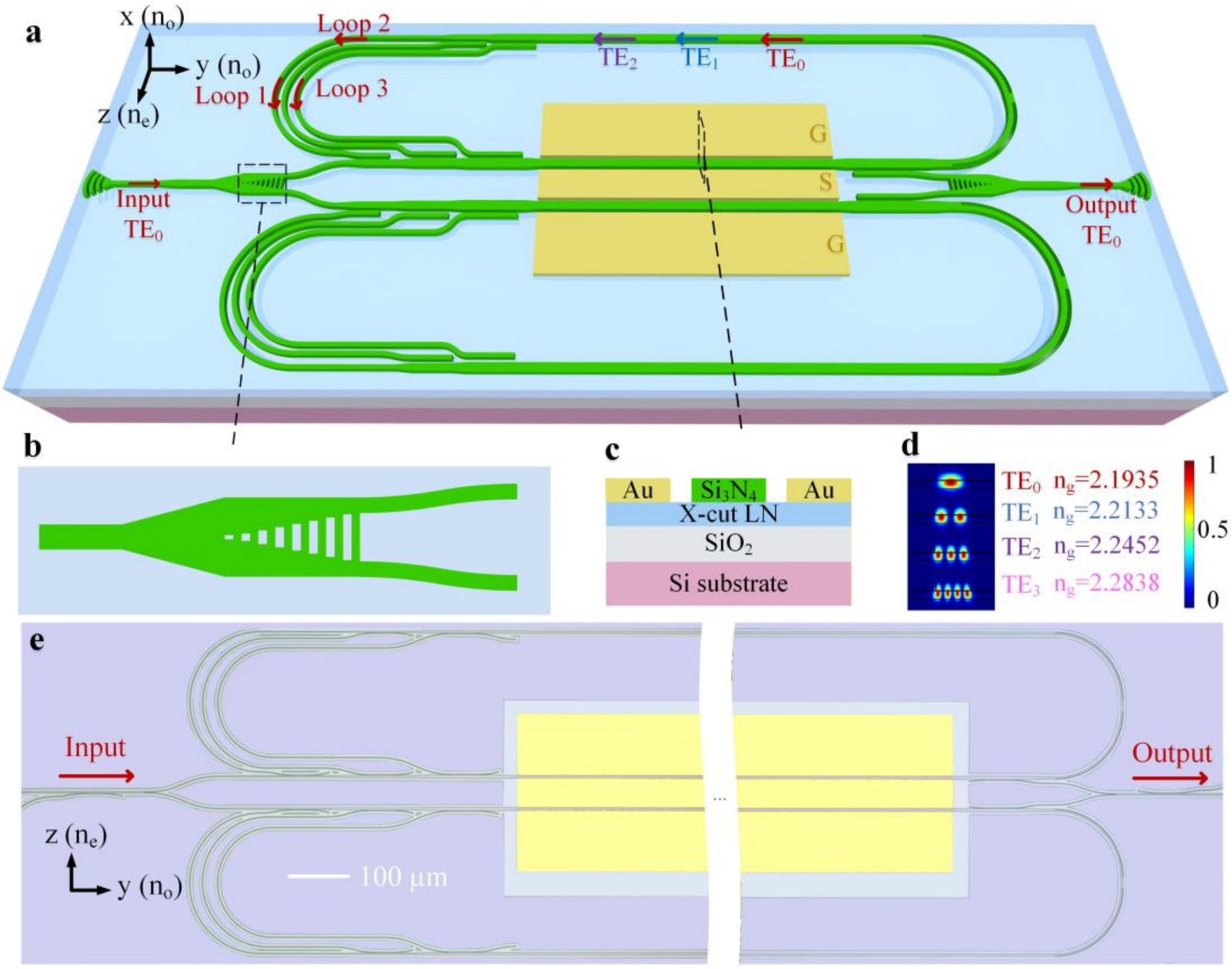

**Figure 4.** MZM on TFLN. **a** Schematic illustration of the MZM architecture. **b** Structural design of the SWG assisted Y-branch. **c** Transverse cross section of the modulation region. **d** Optical field intensity distributions of the supported modes in the multimode waveguide within the modulation region. **e** Optical microscope image of the fabricated MZM.

We next evaluated the EO bandwidth to assess the high frequency performance of the modulator. The experimental setup is shown in Figure 5a. After calibrating out the transmission losses introduced by the RF probes and transmission lines, the extracted 3 dB resonant EO bandwidth exceeds 110 GHz, as shown in Figure 5b. The simulated response agrees well with the experimental results, except for the absence of small sidelobes adjacent to the main resonant peak (see Supplementary Note S5 for details). We attribute this discrepancy to the combined amplitude and phase responses associated with spectral fluctuations induced by crosstalk between the mode converters and the SMWBs. Notably, increasing the number of recirculation loops does not degrade the EO bandwidth of the NRM, in clear contrast to conventional zero-loop modulators with an equivalent electrode length, as discussed in Supplementary Note S6. The dominant limitation on the EO bandwidth originates from the effective index mismatch between the RF and optical signals. The effective index of the RF signal is approximately 2.1, whereas the optical group index is around 2.23. Over the driven RF frequency range from 10 MHz to 110 GHz, four modulation periods are experimentally observed with a frequency step of 10 MHz. The fifth theoretical modulation period, centered at 112 GHz, could not be accessed due to the bandwidth limit of the measurement setup. The measured modulation period is 22.45 GHz, which closely matches the simulated value of 22.39 GHz. The half wave voltage $V_\pi$ of the MZM is measured to be 2.35 V at 10 kHz, corresponding to a $V_\pi$ of 4.7 V for the equivalent PM at the same frequency.

Figures 5c and 5d present the $V_\pi$ values extracted from vector network analyzer (VNA) and optical spectrum analyzer (OSA) measurements at high frequencies[23]. Figure 5d shows a low-$V_\pi$ bandwidth of 4.79 GHz around the first modulation period. Figure 5e summarizes the measured minimum $V_\pi$ and the corresponding low-$V_\pi$ bandwidth for each modulation period. The minimum $V_\pi$ per modulation period increases gradually with RF frequency, from 5.80 V to 7.16 V, with measured values of 5.80 V, 6.23 V, 6.94 V, and 7.16 V. This trend is consistent with the overall 3 dB EO bandwidth of approximately 110 GHz. Meanwhile, the low-$V_\pi$ bandwidth for each modulation period ranges from 4.70 GHz to 5.15 GHz, with measured values of 4.84 GHz, 4.79 GHz, 4.70 GHz, and 5.15 GHz. These values are about two times larger than those reported for previous RMs, which we attribute to the substantially reduced loop length enabled by the proposed NRM architecture. The RF frequency ranges over which the EO response remains within 3 dB of its peak are 0 to 2.52 GHz, 20.08 to 24.92 GHz, 42.52 to 46.75 GHz, 65.72 to 68.93 GHz, and 87.99 to 91.47 GHz.

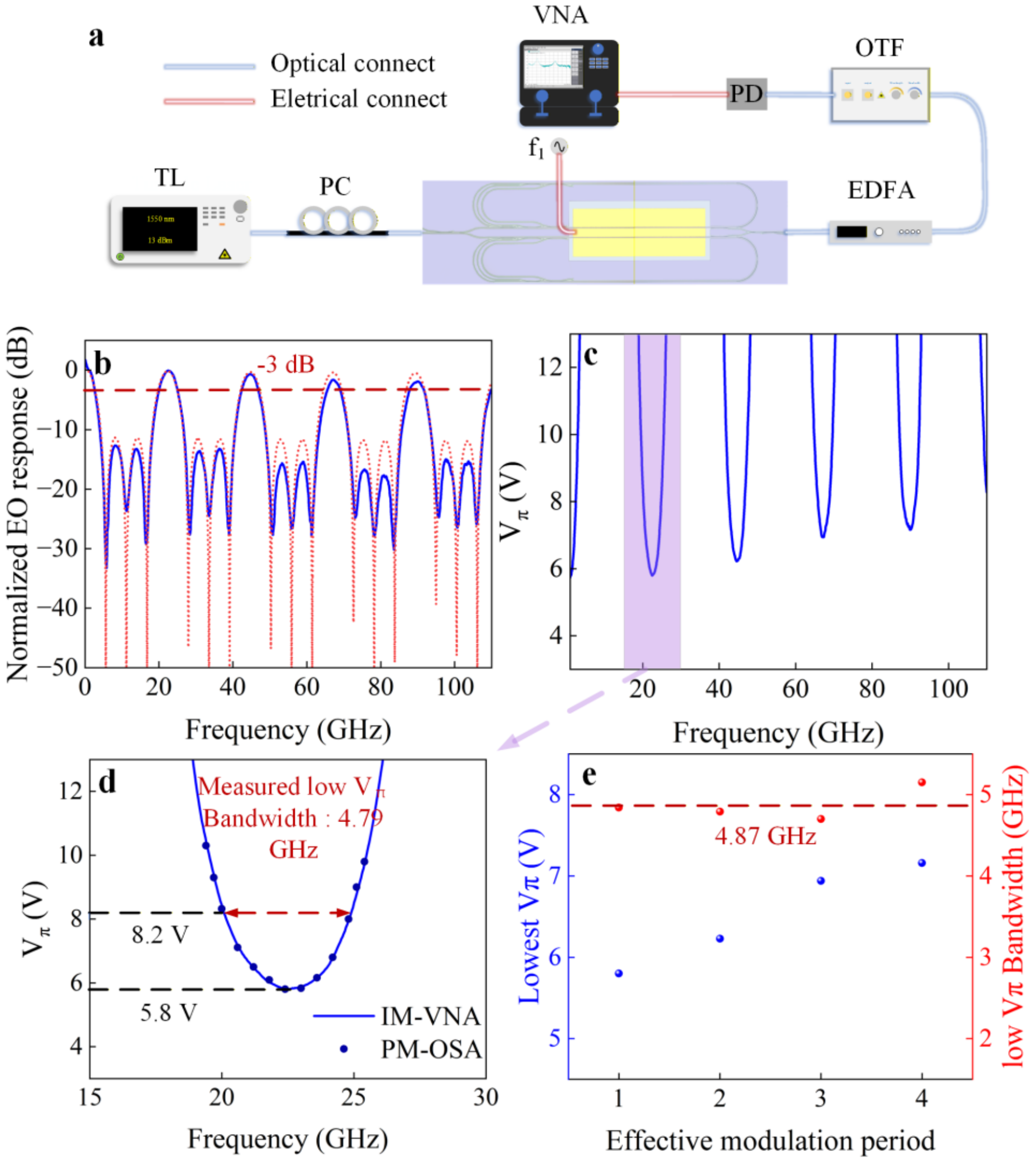

**Figure 5.** EO bandwidth of the MZM and high frequency $V_\pi$ characterization of the PM. **a** Experimental setup for EO response and $V_\pi$ measurements. **b** Measured (solid blue) and simulated (dotted red) EO response from 10 MHz to 110 GHz. **c** PM $V_\pi$ measured from 1 GHz to 110 GHz using the optical method (OSA). **d** PM $V_\pi$ around the first modulation period, measured by the optical method (light blue, OSA) and the electrical method (dark blue dots, VNA). **e** Minimum PM $V_\pi$ and corresponding low-$V_\pi$ bandwidth for each modulation period. TL, tunable laser; PC, polarization controller; VNA, vector network analyzer; EDFA, erbium doped fiber amplifier; OTF, optical tunable filter; PD, photodetector; OSA, optical spectrum analyzer.

### 2.4 Ultra-broadband 4-tap rectangular-window FIR filter

By leveraging the fabricated MZM based on the NRM structure, we establish a new approach for implementing rectangular window FIR filters. Figure 6a and b compare the architectures of a conventional FIR filter and the proposed design. In the proposed four tap FIR filter, the splitter, delay line, amplitude controller, phase controller, and combiner used in a conventional four tap architecture are replaced by six recirculation loops, leading to a substantial improvement in system compactness. The fundamental difference between the proposed approach and conventional schemes lies in converting optical interference among signals propagating along different optical paths at the same time into electrical interference of the same optical path signal at different time instances. Figure 6c compares the results obtained from the proposed simulation approach (red solid line), the conventional calculation method (black dotted line), and experimental measurements (blue solid line). Excellent agreement is observed between our simulation results and those of the conventional method, as detailed in Supplementary Note S7. The sidelobes measured experimentally are more suppressed than those predicted by simulation. This discrepancy is likely caused by perturbations in the spectral response arising from crosstalk between the mode converters and the SMWBs. A comparison with previously reported FIR filters, summarized in Table 1, highlights the significant advances in operating bandwidth, system integration, and compactness achieved in this work. These improvements can be attributed primarily to the adoption of the NRM structure.

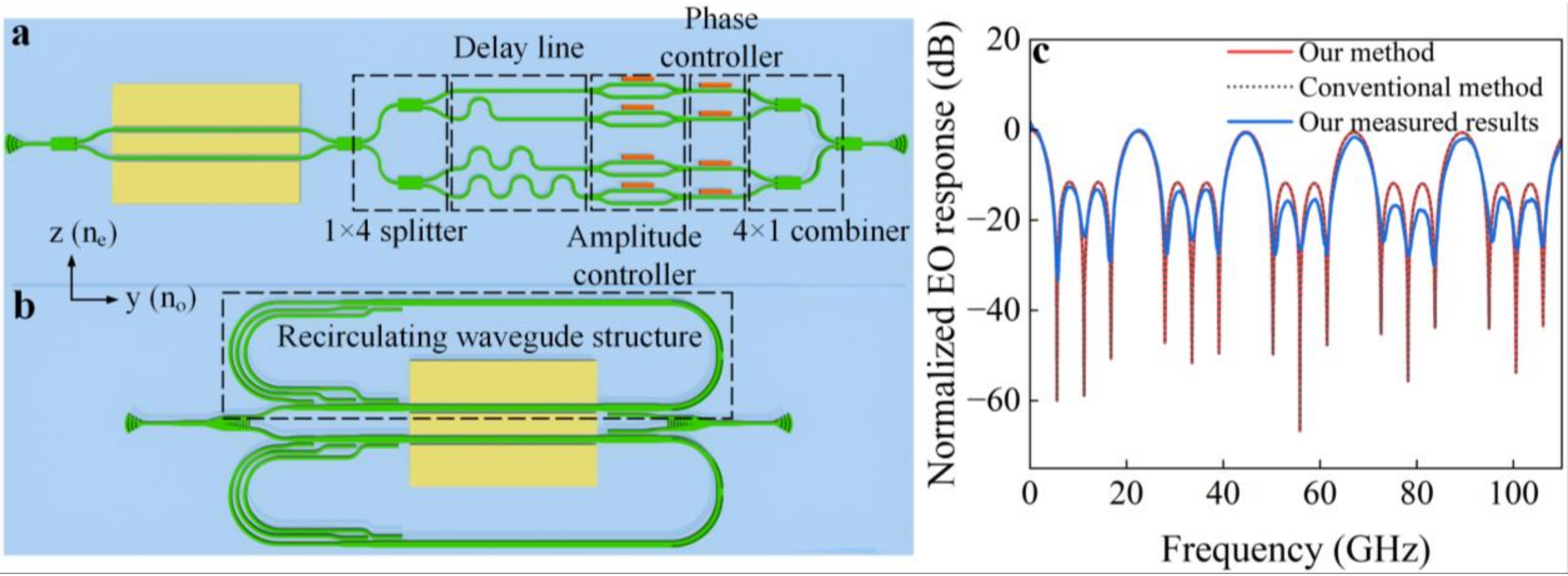

**Figure 6.** The structure comparison of the novel and conventional microwave photonics FIR filter. **a** The conventional FIR filter, including 1×4 optical splitter, delay lines, amplitude control module, phase control module and 4×1 optical combiner. **b** The novel FIR filter, including novel recirculating waveguide structure. **c** Simulated conventional method (black line) and our method (red line) and our measured results (blue line) of the 4-tap FIR filter.

**Table 1. Comparison of FIR filters reported in this work and prior studies.**

| Ref. | Integrated device | Working BW(Hz) | Tap | FSR (Hz) | 3dB BW (Hz) | Q | Shape factor |
|---|---|---|---|---|---|---|---|
| 2011[24] | Delay lines | 0~2.2G | 2 | 93.6M | ~52M | 1.8 | 1.66 |
| 2012[25] | Photonic crystal | 0~40G | 4 | 40G | 8G | 5 | 1.62 |
| 2013[26] | Discrete | 0~31G | 4 | 216.8M | 53M | 4.09 | 1.66 |
| 2016[27] | Discrete | 0~9G | 2 | 3.116G | ~1.57G | 1.98 | 1.55 |
| 2017[28] | Discrete | 0~20G | 8 | 16.9G | ~5.08G | 3.326 | 2.46 |
| 2017[29] | Discrete | 0~7G | 55 | 5.25G | 0.8G | 6.563 | 1.99 |
| 2017[30] | Discrete | 0~40G | 2 | 11.729G | 6.45G | 1.818 | 1.48 |
| 2019[31] | Discrete | 0~18G | 2 | 3.213G | ~1.713G | 1.876 | 1.56 |
| 2022[32] | Laser, comb, delay lines | 0~33.6G | 8 | 7.2G | ~0.93G | 7.74 | 1.76 |
| 2024[33] | MZM, delay lines | 0~27G | 8 | ~17G | 2.04G | 8.33 | 1.68 |
| This work | MZM, delay lines | 0~110G | 4 | 22.39G | 4.87G | 4.598 | 1.72 |

Note: FSR: free spectral range. BW: bandwidth. Q: quality factor.

### 2.5 Ultra-high modulation efficiency frequency combs

Based on the NRM architecture, we fabricated a PM1 that incorporates a single 2mm modulation region, corresponding to one arm of the MZM described above. To further enhance the modulation efficiency, we also implemented a PM2 with two cascaded 1 mm modulation regions, as illustrated in Figure 7a. The operating principle of PM2 is as follows. The input optical signal first propagates through the upper NRM structure and experiences four modulation cycles induced by the upper modulation region. The light is then extracted as the $TE_0$ mode through the single mode waveguide loop 4. The length of loop 4 is precisely engineered to satisfy the 33.39 GHz modulation period dictated by the 1 mm NRM structure. The light is subsequently routed back to the lower modulation region via loop 4 and undergoes an additional four modulation cycles. Finally, the modulated light is coupled out as the $TE_0$ mode, passes through a waveguide crossing, and is delivered to the output port. In total, the optical signal experiences eight modulation cycles across the two modulation regions.

For the parallel modulator configuration, the EO bandwidth was characterized using the same experimental setup described earlier (cf. Figure 5a, rearranged here as Figure 7b). The EO bandwidth analysis is shown in Figure 7c, with additional details provided in Supplementary Note S5. After correcting for probe and transmission line losses, the extracted 3 dB resonant EO bandwidth exceeds 110 GHz. The simulated results are in good agreement with the experimental data, and the remaining bandwidth limitation is primarily attributed to the effective index mismatch between the RF mode (effective index ≈ 2.1) and the optical mode (group index ≈ 2.23). Within the driven RF frequency range from 10 MHz to 110 GHz, sampled with a 10 MHz step, three modulation periods were clearly observed. The fourth theoretically predicted period, centered at 132 GHz, could not be measured owing to experimental limitations. The experimentally extracted modulation period is 33.47 GHz, which closely matches the simulated value of 33.39 GHz.

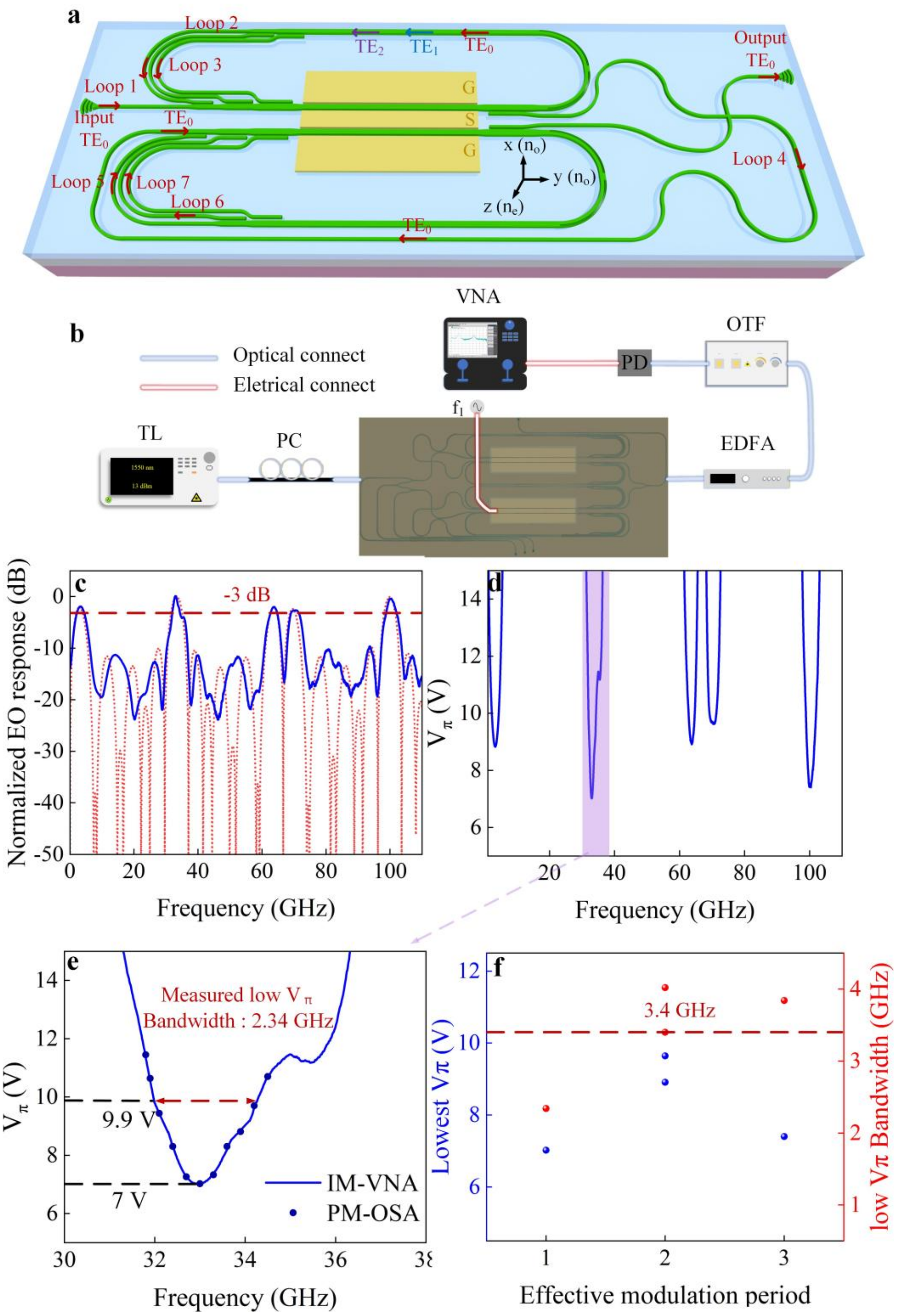

**Figure 7.** The 1 mm PM type II on thin-film lithium niobate. **a** Illustrative drawing outlining the structure of the PM2. **b** Experimental setup. **c** Measured (blue solid line) and simulated (red dotted curve) EO bandwidth from 10 MHz to 110 GHz. **d** Measured PM $V_\pi$ from 1 GHz to 110 GHz. **e.** Measured PM $V_\pi$ near the first modulation period. (light blue line from optical method OSA and dark blue dots from electrical method VNA). **f** Measured lowest $V_\pi$ and low-$V_\pi$ bandwidth in each modulation period. TL: tunable laser, PC: polarization controller, VNA: vector network analyzer, EDFA: erbium-doped fiber amplifier, OTF: optical tunable filter, PD: photodetector, OSA: optical spectrum analyzer.

The half wave voltage $V_\pi$ extracted from VNA and OSA measurements is presented in Figure

7d and Figure 7e, respectively. A low-$V_\pi$ bandwidth of 2.34 GHz is obtained near the first modulation period, as shown in Figure 7e. Figure 7f summarizes the lowest $V_\pi$ value and the corresponding low $V_\pi$ bandwidth for each modulation period. The minimum $V_\pi$ per period exhibits a gradual increase with RF frequency, ranging from 7.00 V to 9.64 V (measured values of 7.00, 8.91, 9.64, and 7.40 V), which is consistent with the measured 3 dB EO bandwidth of approximately 110 GHz. Meanwhile, the low $V_\pi$ bandwidth per period varies between 2.34 GHz and 4.02 GHz, with measured values of 2.34, 3.40, 4.02, and 3.84 GHz. The RF frequency ranges over which the EO response remains within −3 dB are 2.19–4.37 GHz, 31.91–34.30 GHz, 62.77–64.68 GHz, 69.19–71.35 GHz, and 98.87–101.99 GHz.

The experimental setup for E-O comb generation is illustrated in Figure 8a. Figure 8b shows the measured comb spectra of devices PM1 and PM2. In Figure 8b i–iii, PM1 exhibits a progressive increase in the number of comb lines as the driving power rises at a fixed frequency of 22.4 GHz, yielding 9 lines at 18 dBm, 14 lines at 24 dBm, and 27 lines at 31 dBm. Similarly, Figure 8c i–iii present the comb spectra of PM2 under a fixed frequency of 33 GHz, where the number of comb lines increases from 7 at 21 dBm to 11 at 27 dBm and reaches 20 at 33 dBm. The generated comb lines are defined as spectral components whose power exceeds the noise floor. A comparison with previously reported studies based on the TFLN platform, summarized in Table 2, highlights the pronounced reduction in $V_\pi L$ and the simultaneous enhancement of low-$V_\pi$ bandwidth achieved in this work. These improvements are enabled by the NRM structure. Further reduction of $V_\pi L$ can be realized by decreasing the signal ground electrode spacing or by introducing additional modes into the modulation region.

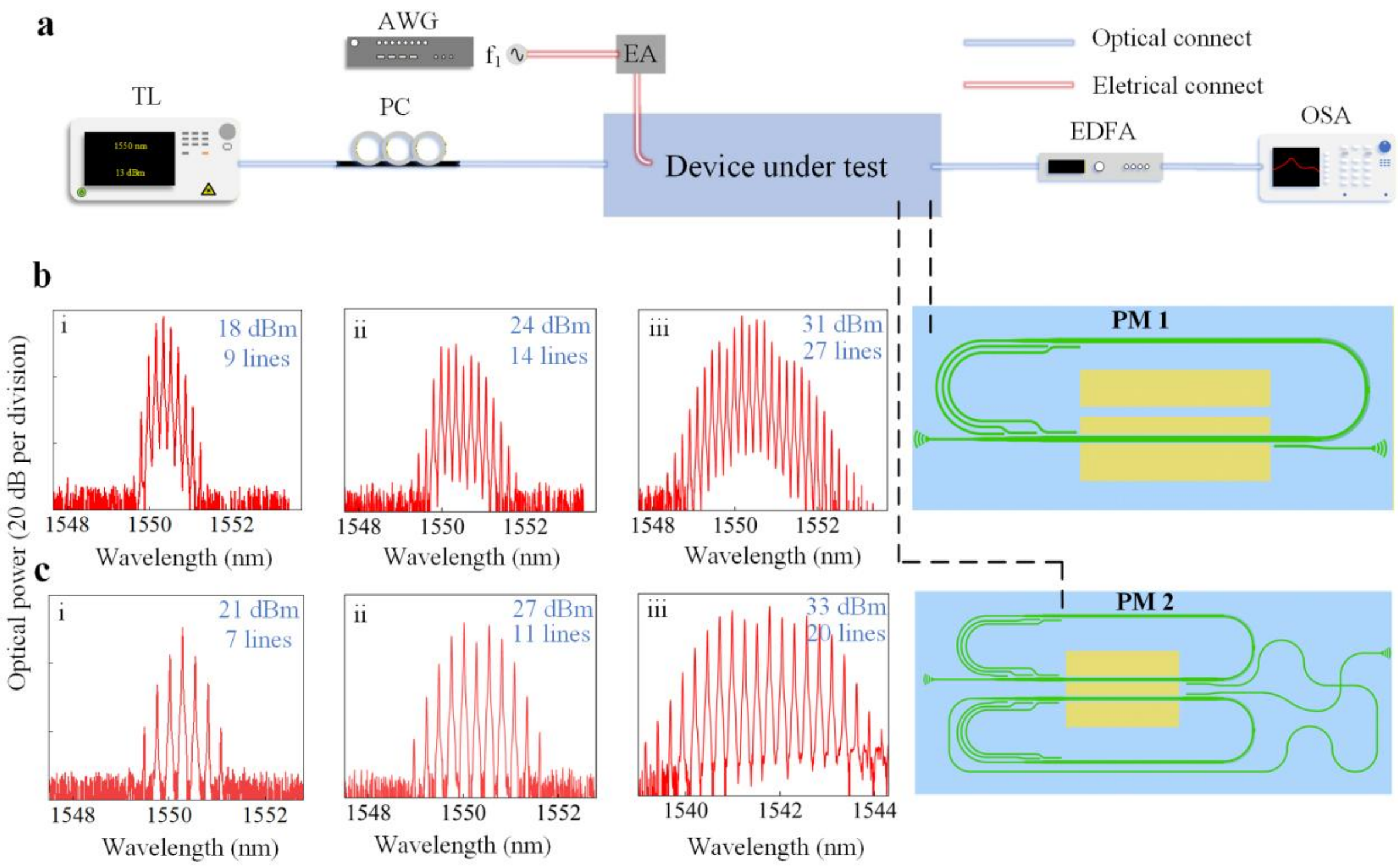

**Figure 8.** Measurement of ultra-high-modulation-efficiency E-O frequency combs. **a** Experimental setup. **b** Measured spectra of PM1 driven at 22.4 GHz with RF powers of (i) 18 dBm, (ii) 24 dBm, and (iii) 31 dBm. **c** Measured spectra of PM2 driven at 33 GHz with RF powers of (i) 21 dBm, (ii) 27 dBm, and (iii) 33 dBm. TL, tunable laser; PC, polarization controller; AWG, arbitrary waveform generator; EA, electrical amplifier; OSA, optical spectrum analyzer. PM1, 2 mm phase modulator (type I); PM2, 1 mm phase modulator (type II).

**Table 2. Comparison of phase modulators reported on TFLN platform.**

| Ref. | Length (cm) | $V_\pi$ (V) | $V_\pi L$= (V•cm) | EO BW (Hz) | Low-$V_\pi$ BW (Hz) |
|---|---|---|---|---|---|
| 2019[13] | 2 | 3.5 @5 GHz | 7 | Not mentioned | / |
| 2022[3] | 2 | ~ 2 @12 GHz | 4 | Not mentioned | 1.4G |
| 2023[16] | 1 | 1.9 @25 GHz | 1.9 | 40G, resonant | 1.4G |
| 2024[34] | 2 | 2.5 @30.135 GHz | 5 | Not mentioned | ~ 1.41G |
| 2024[35] | 3 | 1.45 @~ 11.5 GHz | 4.35 | Not mentioned | ~ 1.23G |
| 2024[36] | 5 | 1.5 @18 GHz | 7.5 | Not mentioned | ~ 1.35G |
| 2024[8] | 1 | 3 @12 GHz | 3 | Not mentioned | ~ 2.63G |
| 2024[37] | 2 | 2 @~ 14.5 GHz | 4 | ~ 38G, resonant | 3G |
| 2025[38] | 0.745 | 3.92 @~ 14.5 GHz | 2.92 | >67G, resonant | 3.3G |
| This work, MZM | 0.2 | 2.35 @10 kHz | 0.47 | >110G, resonant | 4.87G |
| This work, PM1 | 0.2 | 5.8 @22.4 GHz | 1.16 | >110G, resonant | 4.87G |
| This work, | 0.1 | 7 @33 GHz | 0.7 | >110G, resonant | 3.40G |

| PM2 | | | | | |
|---|---|---|---|---|---|

Note: EO BW: electro-optic bandwidth.

## 3. Discussion

In conventional RMs, strict loop length matching imposes a strong coupling between extra device length, electrode length, and the number of supported modes. With a fixed metal electrode length, increasing the number of modes to enhance modulation efficiency leads to a proportional increase in extra waveguide length, and this penalty becomes more severe as the mode count grows. Similarly, for a fixed number of modes, longer electrodes require proportionally longer loop-back sections. In addition, the loop length must often be extended by several multiples to satisfy phase matching, which significantly degrades the low-$V_\pi$ bandwidth. These intrinsic trade-offs severely limit the practical scalability of conventional RMs. We demonstrate a novel non-resonant recirculating phase modulator realized through the synergistic integration of two key elements. The first is a carefully engineered multi-pass phase modulation configuration that enhances modulation efficiency by up to a factor of four. The second is a set of low-loss, low-crosstalk $TE_0/TE_1$, $TE_0/TE_2$, and $TE_0/TE_3$ mode converters combined with SMWBs that support the $TE_0$, $TE_1$, and $TE_2$ modes. This architecture is implemented on the TFLN platform and is specifically designed to overcome the loop length matching constraint that limits conventional RMs, while simultaneously improving compactness and expanding the low-$V_\pi$ bandwidth.

To validate the advantage of the proposed architecture, we fabricated 1 to 4 loop RM variants. For the 3-mode and 4-mode devices with 3 mm electrodes, the measured extra lengths are only 0.730 mm and 0.889 mm, corresponding to effective modulation length percentages of 80.44% and 77.14%, respectively. Importantly, the extra length shows negligible dependence on either the mode number or the electrode length. As the electrode length increases, the effective modulation length percentage improves markedly, while the extra length remains nearly constant. Low-$V_\pi$ bandwidth measurements further corroborate this advantage. A single-modulation-region RM with a 2 mm electrode exhibits a 4.87 GHz low-$V_\pi$ bandwidth, compared with only 1.27 GHz for a conventional RM with the same electrode length and operating frequency. Likewise, a dual-modulation-region RM with a 1 mm electrode achieves a 3.4 GHz low-$V_\pi$ bandwidth, substantially outperforming the 1.05 GHz bandwidth of its conventional counterpart.

Tapped delay-line filters, also known as FIR filters, constitute the most established architecture for microwave photonic filters[39,40]. Devices with a finite number of taps fall into the class of FIR filters. Most FIR filters reported to date rely on discrete modulators and separated delay lines, resulting in bulky, inefficient, and table-top scale systems with limited operating bandwidths[41,42]. As a representative application of the proposed NRM structure, we experimentally realize a MZM configured as a 4-tap rectangular-window FIR filter. This implementation achieves both the most compact FIR filter footprint reported to date, measuring only 2.889 × 0.58 $mm^2$, and the widest FIR filter bandwidth, reaching 110 GHz. Such performance highlights strong potential for future terahertz applications, including high-speed wireless communication for 6G systems[43], THz radar[44], and THz time-domain spectroscopy[45].

Conventional on-chip recirculating phase modulators on the thin-film lithium niobate platform are still limited by relatively large $V_\pi L$ values and narrow low-$V_\pi$ bandwidths. Using the proposed NRM structure, we further fabricate two devices, PM1 and PM2, featuring 4-mode and 8-mode configurations. To the best of our knowledge, these devices jointly achieve the broadest low-$V_\pi$ bandwidths of 4.87 GHz and 3.40 GHz, a resonant EO bandwidth exceeding 110 GHz, and the lowest $V_\pi L$ values of 1.16 V·cm and 0.7 V·cm among comparable implementations. Under microwave drive powers of 31 dBm and 33 dBm, PM1 and PM2 generate 27 and 20 comb lines, respectively. These frequency comb generators offer a promising multiwavelength source for future optical computing[46], high-capacity optical interconnects[47], and optical clocks[48].

Despite the demonstrated ultrabroad operating bandwidth and compact footprint, the response of the fabricated FIR filter is currently fixed because the loop length determines a static delay. Different application scenarios require distinct and tunable bandwidth specifications. This limitation can be addressed by integrating an on-chip tunable time delay line into the loop-back waveguide. By precisely adjusting the optical delay, the FIR filter response can be tuned in a flexible manner. Incorporation of an on-chip photodetector is also feasible and would further reduce system size while lowering power consumption for advanced miniaturized applications.

For the frequency comb generators, increasing the number of comb lines can be achieved by extending the electrode length to reduce $V_\pi$. However, this approach inevitably degrades the low-$V_\pi$ bandwidth due to the increased loop length. Increasing the number of supported modes offers another route to higher modulation efficiency, but our simulations indicate that additional mode loops introduce larger delays, which again reduce the low-$V_\pi$ bandwidth. PM1 and PM2 can be integrated with a MZM to generate flat-top broadband EO combs under low RF drive power. Dispersive elements placed downstream can further enable ultrashort pulse generation [3].

Full miniaturization and self-sustained operation represent key goals for next-generation FIR filters and EO comb generators. Progress toward these goals will rely on the development of highly integrated on-chip light sources. Although current systems using external lasers have successfully validated the proposed architectures[49,50], future advances are expected to exploit hybrid integration of high-performance quantum dot lasers or heterogeneous integration approaches. These strategies promise compact, low-power photonic terminals with significantly improved practicality and compatibility, aligned with the evolving demands of FIR filter and EO comb generator technologies.

## 4. Methods

**Fabrication of the Photonic Chip**

The proposed NRM modulators were fabricated via a sequence of electron beam lithography (EBL) and inductively coupled plasma (ICP) etching processes. Specifically, *X*-cut TFLN wafers supplied by NanoLN were selected as the substrate. This choice was motivated by the desire to leverage the prominent electro-optical tensor component $\gamma_{33}$ of LN, which serves to boost the operational performance of modulators with Y-directional propagation. Prior to device patterning, a thin film of $Si_3N_4$ was deposited onto the TFLN surface using a reactive sputtering technique. Following this, EBL was employed for resist patterning, and ICP etching was utilized to define the waveguide and grating coupler architectures with high precision. Finally, the electrode structures were fabricated through a series of refined processes, encompassing direct laser writing, electron beam evaporation, and a subsequent lift-off step to achieve the desired electrode geometry.

**Experimental Setup**

The experimental setup for the NRM and OFC measurement is utilizing a continuous wave from a tunable laser (TL) which is routed through a polarization controller (PC), amplified by an erbium-doped fiber amplifier (EDFA), and coupled onto the chip via an integrated grating coupler. Microwave signals from an arbitrary waveform generator (AWG) are amplified by an electrical amplifier (EA) and delivered to the RM through a high-speed 50 Ω ground-signal-ground (GSG) probe, with the electrode output terminated by 50 Ω impedance-matching terminators. Chip-emitted light is captured by an optical spectrum analyzer (OSA) for spectral analysis. The applied power is calibrated by excluding the losses from the high-frequency cables and probe.

In the EO bandwidth experiment to characterize the MZM, PM1 and PM2 high-frequency performance, a vector network analyzer (VNA) served as the RF signal source, delivering signals to the modulator via a 50 Ω GSG probe. The VNA was also used to capture the output optical signal for EO bandwidth analysis. After accounting for transmission losses from the probe and transmission line, the derived results are captured by VNA.

## 5. Data Availability Statement

All the data supporting the findings in this study are available in the paper and Supplementary Information. Additional data related to this paper are available from the corresponding authors upon request.

## Acknowledgements

This work was supported by National Natural Science Foundation of China (NSFC)(W2411059, 62405125, 62205135), Gansu Provincial Science and Technology Major Special Project (23ZDGE001, 25ZDWA001, 25ZDGA005), Key Research and Development of Gansu Province (24YFGA007), China Postdoctoral Science Foundation (2025M780782), The Joint Research Fund Project of Gansu Province (25JRRA1126) and Australian Research Council (ARC) (CE230100006).

## Author contributions

All authors contributed extensively to the work presented in this paper. F. W. conceived the study. F. W., L. W, Z. H., B. W., and Y. Z. designed and characterized the NRM chip. P. Z., Y. J., M. Y., and H. X. conducted the novel non-resonant recirculating modulator measurement experiment. P. Z., and F. W. performed the EO bandwidth experiment. M.X.L., A.D., T.G.N., and F. W. performed the optical frequency comb measurement. F. W. conducted theoretical analysis. F. W., M. Y., H. X., G. R., A. M., and Y. T. wrote the manuscript with input from all coauthors. Y. T. supervised the research. All authors participated in data analysis and discussion.

## Competing Interests Statement

The authors declare no competing interests.

**Table 1. Comparison of FIR filters reported in this work and prior studies.**

| Ref. | Integrated device | Working BW(Hz) | Tap | FSR (Hz) | 3dB BW (Hz) | Q | Shape factor |
|---|---|---|---|---|---|---|---|
| 2011[24] | Delay lines | 0~2.2G | 2 | 93.6M | ~52M | 1.8 | 1.66 |
| 2012[25] | Photonic crystal | 0~40G | 4 | 40G | 8G | 5 | 1.62 |
| 2013[26] | Discrete | 0~31G | 4 | 216.8M | 53M | 4.09 | 1.66 |
| 2016[27] | Discrete | 0~9G | 2 | 3.116G | ~1.57G | 1.98 | 1.55 |
| 2017[28] | Discrete | 0~20G | 8 | 16.9G | ~5.08G | 3.326 | 2.46 |
| 2017[29] | Discrete | 0~7G | 55 | 5.25G | 0.8G | 6.563 | 1.99 |
| 2017[30] | Discrete | 0~40G | 2 | 11.729G | 6.45G | 1.818 | 1.48 |
| 2019[31] | Discrete | 0~18G | 2 | 3.213G | ~1.713G | 1.876 | 1.56 |
| 2022[32] | Laser, comb, delay lines | 0~33.6G | 8 | 7.2G | ~0.93G | 7.74 | 1.76 |
| 2024[33] | MZM, delay lines | 0~27G | 8 | ~17G | 2.04G | 8.33 | 1.68 |
| This work | MZM, delay lines | 0~110G | 4 | 22.39G | 4.87G | 4.598 | 1.72 |

Note: FSR: free spectral range. BW: bandwidth. Q: quality factor.

**Table 2. Comparison of phase modulators reported on TFLN platform.**

| Ref. | Length (cm) | $V_\pi$ (V) | $V_\pi L$= (V•cm) | EO BW (Hz) | Low-$V_\pi$ BW (Hz) |
|---|---|---|---|---|---|
| 2019[13] | 2 | 3.5V @5 GHz | 7 | Not mentioned | / |
| 2022[3] | 2 | ~ 2 V @12 GHz | 4 | Not mentioned | 1.4G |
| 2023[16] | 1 | 1.9 V @25 GHz | 1.9 | 40G, resonant | 1.4G |
| 2024[34] | 2 | 2.5V @30.135 GHz | 5 | Not mentioned | ~ 1.41G |
| 2024[35] | 3 | 1.45 V @~ 11.5 GHz | 4.35 | Not mentioned | ~ 1.23G |
| 2024[36] | 5 | 1.5 V @18 GHz | 7.5 | Not mentioned | ~ 1.35G |
| 2024[8] | 1 | 3 V @12 GHz | 3 | Not mentioned | ~ 2.63G |
| 2024[37] | 2 | 2 V @~ 14.5 GHz | 4 | ~ 38G, resonant | 3G |
| 2025[38] | 0.745 | 3.92 V @~ 14.5 GHz | 2.92 | >67G, resonant | 3.3G |
| This work, MZM | 0.2 | 2.35 V @10 kHz | 0.47 | >110G, resonant | 4.87G |
| This work, PM1 | 0.2 | 5.8 V @22.4 GHz | 1.16 | >110G, resonant | 4.87G |
| This work, PM2 | 0.1 | 7 V @33 GHz | 0.7 | >110G, resonant | 3.40G |

Note: EO BW: electro-optic bandwidth.

**Figure 1.** **a** Schematic overview of the proposed NRMs. **b** Structural layouts of the $TE_0/TE_1$, $TE_0/TE_2$, and $TE_0/TE_3$ mode converters. **c** Cross-sectional view of the modulation region. **d** Cross section of the SMWBs. **e** Simulated optical field intensity distributions in the multimode waveguide within the modulation region. **f** Overview of the slotted multimode waveguide bends. **g** Optical micrograph of the fabricated three-loop NRM. **h** Fabricated SMWBs supporting $TE_0$, $TE_1$, and $TE_2$ modes. **i** Fabricated traveling-wave electrodes and the input multimode waveguide. **j** Fabricated $TE_0/TE_3$ mode converter. **k** Fabricated $TE_0/TE_2$ mode converter. **l** Fabricated $TE_0/TE_1$ mode converter.

---

**Figure 2.** Measurement of the proposed NRM. **a** Experimental setup used for modulation characterization. **b** Measured optical spectra obtained with adjusted RF drive voltages to equalize the modulation index: i 500 mV applied to PM, ii 250 mV to RM1, iii 166.7 mV to RM2, and iv 125 mV to RM3. v The corresponding extracted modulation indices. **c** Measured optical spectra under a fixed RF drive voltage of 700 mV applied to i PM, ii RM1, iii RM2, and iv RM3. v The corresponding extracted modulation indices. TL, tunable laser; PC, polarization controller; AWG, arbitrary waveform generator; EA, electrical amplifier; EDFA, erbium doped fiber amplifier; OSA, optical spectrum analyzer; PM, phase modulator; RM1, recirculating modulator with one loop; RM2, recirculating modulator with two loops; RM3, recirculating modulator with three loops.

---

**Figure 3.** Modulation efficiency, footprint, and low-$V_\pi$ bandwidth analysis of NRMs. **a** Modulation index of PM, RM1, RM2, and RM3 as a function of the applied RF power. **b** Modulation index measured across the C band as a function of operational wavelength for different modulators. **c** Theoretical analysis of the effective modulation length percentage for the proposed and conventional RM structures. The dotted orange and blue curves indicate the simulated results for RM4, which was not fabricated. **d** Theoretical analysis of the extra length for the proposed and conventional RM structures. The dotted orange and blue curves correspond to the simulated RM4. **e** Theoretical analysis together with representative literature data of the low-$V_\pi$ bandwidth for the RM with one loop. **f** Theoretical analysis of the low-$V_\pi$ bandwidth for the proposed and conventional RM structures with two loops. **g** Theoretical analysis and our experimental results for PM1 and PM2, illustrating the low-$V_\pi$ bandwidth of the proposed and conventional RM structures with three loops. PM, phase modulator; RM, recirculating modulator; RM1, recirculating modulator with one loop; RM2, recirculating modulator with two loops; RM3, recirculating modulator with three loops; ORM1, conventional recirculating modulator with one loop; ORM2, conventional recirculating modulator with two loops; ORM3, conventional recirculating modulator with three loops; NRM2, novel recirculating modulator with two loops; NRM3, novel recirculating modulator with three loops; PM1, 2 mm PM type I; PM2, 1 mm PM type II.

---

**Figure 4.** MZM on TFLN. **a** Schematic illustration of the MZM architecture. **b** Structural design of the SWG assisted Y-branch. **c** Transverse cross section of the modulation region. **d** Optical field intensity distributions of the supported modes in the multimode waveguide within the modulation region. **e** Optical microscope image of the fabricated MZM.

**Figure 5.** EO bandwidth of the MZM and high frequency $V_\pi$ characterization of the PM. **a** Experimental setup for EO response and $V_\pi$ measurements. **b** Measured (solid blue) and simulated (dotted red) EO response from 10 MHz to 110 GHz. **c** PM $V_\pi$ measured from 1 GHz to 110 GHz using the optical method (OSA). **d** PM $V_\pi$ around the first modulation period, measured by the optical method (light blue, OSA) and the electrical method (dark blue dots, VNA). **e** Minimum PM $V_\pi$ and corresponding low-$V_\pi$ bandwidth for each modulation period. TL, tunable laser; PC, polarization controller; VNA, vector network analyzer; EDFA, erbium doped fiber amplifier; OTF, optical tunable filter; PD, photodetector; OSA, optical spectrum analyzer.

**Figure 6.** The structure comparison of the novel and conventional microwave photonics FIR filter. **a** The conventional FIR filter, including 1×4 optical splitter, delay lines, amplitude control module, phase control module and 4×1 optical combiner. **b** The novel FIR filter, including novel recirculating waveguide structure. **c** Simulated conventional method (black line) and our method (red line) and our measured results (blue line) of the 4-tap FIR filter.

**Figure 7.** The 1 mm PM type II on thin-film lithium niobate. **a** Illustrative drawing outlining the structure of the PM2. **b** Experimental setup. **c** Measured (blue solid line) and simulated (red dotted curve) EO bandwidth from 10 MHz to 110 GHz. **d** Measured PM $V_\pi$ from 1 GHz to 110 GHz. **e.** Measured PM $V_\pi$ near the first modulation period. (light blue line from optical method OSA and dark blue dots from electrical method VNA). **f** Measured lowest $V_\pi$ and low-$V_\pi$ bandwidth in each modulation period. TL: tunable laser, PC: polarization controller, VNA: vector network analyzer, EDFA: erbium-doped fiber amplifier, OTF: optical tunable filter, PD: photodetector, OSA: optical spectrum analyzer.

**Figure 8.** Measurement of ultra-high-modulation-efficiency E-O frequency combs. **a** Experimental setup. **b** Measured spectra of PM1 driven at 22.4 GHz with RF powers of (i) 18 dBm, (ii) 24 dBm, and (iii) 31 dBm. **c** Measured spectra of PM2 driven at 33 GHz with RF powers of (i) 21 dBm, (ii) 27 dBm, and (iii) 33 dBm. TL, tunable laser; PC, polarization controller; AWG, arbitrary waveform generator; EA, electrical amplifier; OSA, optical spectrum analyzer. PM1, 2 mm phase modulator (type I); PM2, 1 mm phase modulator (type II).

Supplementary information for

# Ultra-high modulation efficiency and compact thin film lithium niobate non-resonant recirculating phase modulator and its applications

Feiyu Wang[1,†], Liheng Wang[1,†], Mingrui Yuan[1,†], Zhen Han[1], Binjie Wang[1], Yong Zheng[1], Pu Zhang[1], Yongheng Jiang[1], Huifu Xiao[1], Mei Xian Low[2], Aditya Dubey[2], Thach Giang Nguyen[2], Guanghui Ren[2], Arnan Mitchell[2], and Yonghui Tian[1,3*]

[1] School of Physical Science and Technology, Lanzhou University, Lanzhou 730000, Gansu, China

[2] Integrated Photonics and Applications Centre (InPAC), School of Engineering, RMIT University, Melbourne, VIC 3001, Australia

[3] School of Mathematics and Physics, North China Electric Power University, Beijing 102206, China

[†]These authors contributed equally to this work.

Corresponding author: *tianyh@lzu.edu.cn.

**This supplementary information file includes:**

Supplementary Note S1: The operational principle of recirculating modulator.

Supplementary Note S2: The comparison between the conventional and novel RM.

Supplementary Note S3: Details in designing the NRM.

Supplementary Note S4: Performances of the passive components.

Supplementary Note S5: EO response spectrum simulation.

Supplementary Note S6: EO response comparison between the NRMs and the common modulators.

Supplementary Note S7: The novel approach to generating microwave photonics rectangular window finite impulse response (FIR) filter.

Supplementary Table 1: Physical structure parameters of the passive components

Supplementary Table 2: Structures, widths, lengths and group refractive indices for each part in the novel non-resonant recirculating modulator.

Supplementary Table 3: Structures, widths, lengths and group refractive indices for each part in the Mach-Zehnder modulator.

Supplementary Table 4: Structures, widths, lengths and group refractive indices for each part in the high modulation efficiency phase modulator type II.

## Supplementary Note S1: The operational principle of recirculating modulator

The section presents the comprehensive principles underpinning the recirculating modulator (RM). The microwave signal, characterized by a sinusoidal waveform with an amplitude $V_m$, a frequency $\omega_m$, and an original phase $\varphi$, is formulated as follows:

$$V(t) = V_m \sin(\omega_m t + \varphi) \tag{1}$$

The microwave signal is introduced into the signal electrode of the phase modulator, and the resulting optical signal can be formulated as follows:

$$E(t) = E_c e^{j\omega_c t} e^{j\frac{\pi}{V_\pi} V_m \sin(\omega_m t + \varphi)} \tag{2}$$

where $E_c$ and $\omega_c$ are the amplitude and frequency of the optical carrier and $V_\pi$ is the minimum voltage to achieve a 180° phase shift in the optical signal on the phase modulator. As illustrated in Figure 1(a), in the context of a RM, the light undergoes multiple modulation cycles. For $N$ recirculation loops, this process can be mathematically represented as follows[1]:

$$\begin{aligned} E_{out} &= E_c e^{j\omega f_c} e^{j\pi\frac{V_m}{V_{\pi 0}}\sin(\omega_m t+\varphi)} e^{j\pi\frac{V_m}{V_{\pi 1}}\sin(\omega_m (t+\Delta T_1)+\varphi)} \\ &\times e^{j\pi\frac{V_m}{V_{\pi 2}}\sin(\omega_m (t+\Delta T_1+\Delta T_2)+\varphi)} \cdots e^{j\pi\frac{V_m}{V_{\pi N}}\sin(\omega_m (t+\Delta T_1+\Delta T_2+\cdots+\Delta T_N)+\varphi)} \end{aligned} \tag{3}$$

In which $V_{\pi j}$ (for $j$ ranging from $0$ to $N$) represents the $V_\pi$ of the $j$-th recirculation in the form of $TE_j$, and $\Delta T_j$ (for $j$ from $1$ to $N$) signifies the loop's transit duration for light to revisit the modulator. Given a fixed value of d, the $j$-th loop can be divided into d segments based on variations in the waveguide's width and the order changes of the optical field mode within the waveguide which both will induce the changes of the group velocity, with the $i$-th segment being one of these divisions. $\Delta T_j$ is determined by the formula $\Delta T_j = \sum_{i=1}^{d} \Delta L_{ji}/v_{gi}$, where $\Delta L_{ji}$ represents the length of the $i$-th segment within the $j$-th loop, and $v_{gi}$ represents the group velocity of the mode within the $i$-th segment. If it is assumed that $V_{\pi j}$ remains consistent across all modes, then Equation (3) can be reformulated as follows:

$$E(t) = E_c e^{j\omega_c t} e^{jk\frac{\pi}{V_\pi} V_m \sin(\omega_m t + \varphi + \theta)} = E_c e^{j\omega_c t} e^{j\frac{\pi}{V_\pi / k} V_m \sin(\omega_m t + \varphi + \theta)} \tag{4}$$

where $\theta$ represents the phase shift and $k$ represents the modulation enhancement factor of the RM. For the RM with $N$ recirculation loops, as depicted in equation (3) they can be mathematically expressed as:

$$\theta = \arctan\left(\frac{\sin(\omega_m \Delta T_1) + \sin(\omega_m(\Delta T_1 + \Delta T_2)) + \cdots + \sin(\omega_m(\Delta T_1 + \Delta T_2 + \cdots + \Delta T_N))}{1 + \cos(\omega_m \Delta T_1) + \cos(\omega_m(\Delta T_1 + \Delta T_2)) + \cdots + \cos(\omega_m(\Delta T_1 + \Delta T_2 + \cdots + \Delta T_N))}\right)$$

$$k = \sqrt{(1 + \cos(\omega_m \Delta T_1) + \cos(\omega_m(\Delta T_1 + \Delta T_2)) + \cdots + \cos(\omega_m(\Delta T_1 + \Delta T_2 + \cdots + \Delta T_N)))^2 + \cdots \rightarrow \atop \leftarrow \cdots (\sin(\omega_m \Delta T_1) + \sin(\omega_m(\Delta T_1 + \Delta T_2)) + \cdots + \sin(\omega_m(\Delta T_1 + \Delta T_2 + \cdots + \Delta T_N)))^2} \quad (5)$$

As observed from previous equations, the output signal exhibits significant dependence on the modulation enhancement factor, which is governed by the frequency $\omega_m$ and the delay time $\Delta T_j$ associated with each recirculation loop. The peak enhancement factor occurs when $\Delta T_j$ is set to $2n\pi/\omega_m$, where n is an arbitrary integer, which we call in-phase modulation because in this condition the phases of the microwave signals are identical at the output end. In the case of $N$ recirculation loops, the maximum enhancement factor attains a value of $N$+1.

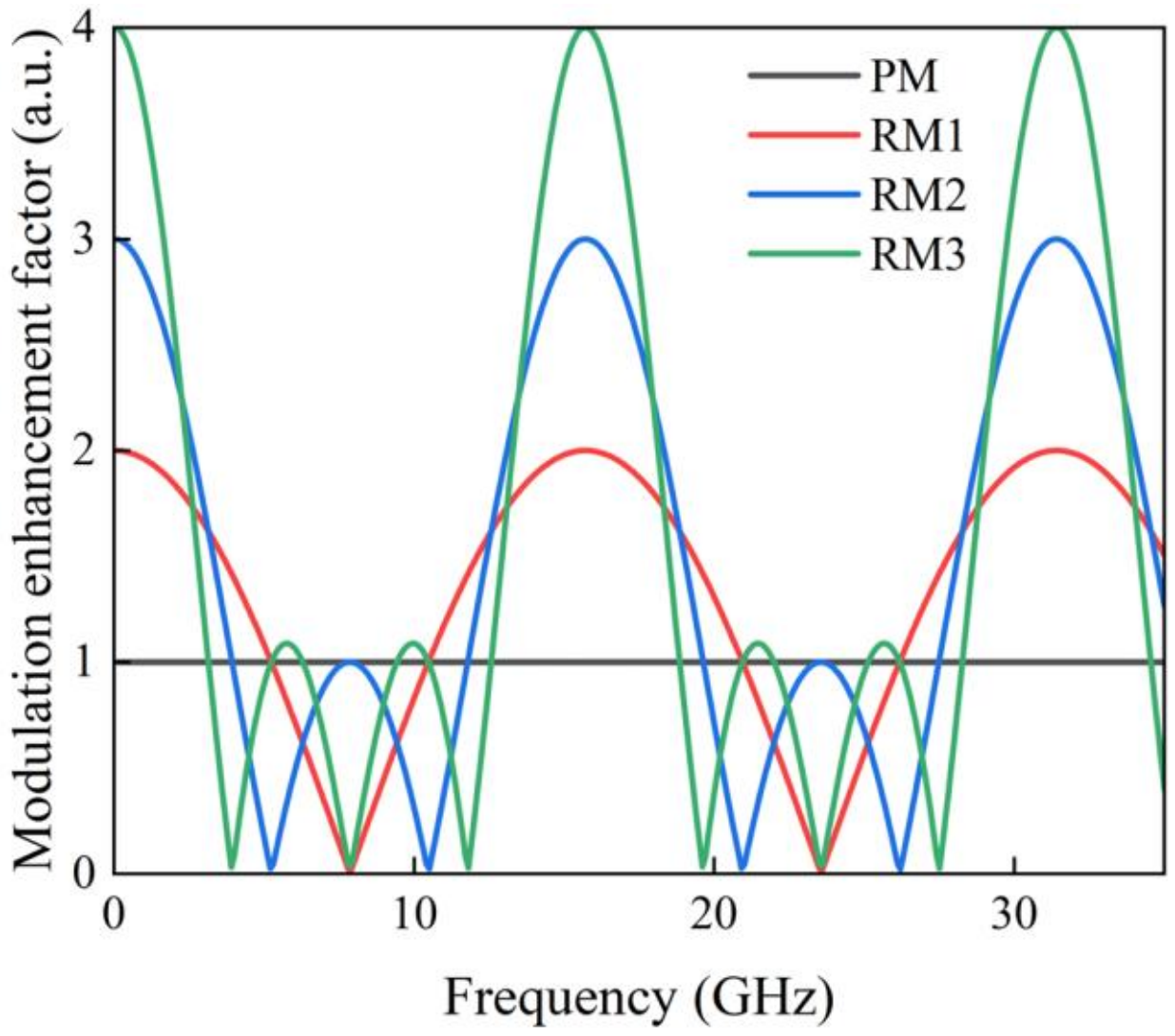

**Figure S1.** The modulation enhancement factors, as they relate to the modulation frequency, are analyzed for PM, RM1, RM2, and RM3. A common characteristic of a typical single-pass PM is the presence of a frequency-independent baseline enhancement factor, denoted as k = 1. PM: phase modulator, RM1: recirculating modulator with one loop, RM2: recirculating modulator with two loops, RM3: recirculating modulator with three loops.

According to equation (5), we conducted simulations to depict the modulation enhancement factor in relation to the frequency, with the results presented in Figure S1.

Upon inspection, it becomes evident that a conventional single-pass phase modulator (PM) maintains a consistent enhancement factor of 1 across all frequencies, under the assumption of optimal performance. In contrast, a recirculating modulator configured with one loop (RM1) exhibits a periodic behavior, achieving an enhancement factor of approximately $k = 2$ at specific frequencies of 0, 16.9, and 33.8 GHz. Notably, the recirculating modulator with two loops (RM2) attains an enhancement factor of nearly $k = 3$, while the recirculating modulator with three loops (RM3) reaches an enhancement factor of approximately $k = 4$ at the corresponding frequencies.

## Supplementary Note S2: The comparison between the conventional and novel RM.

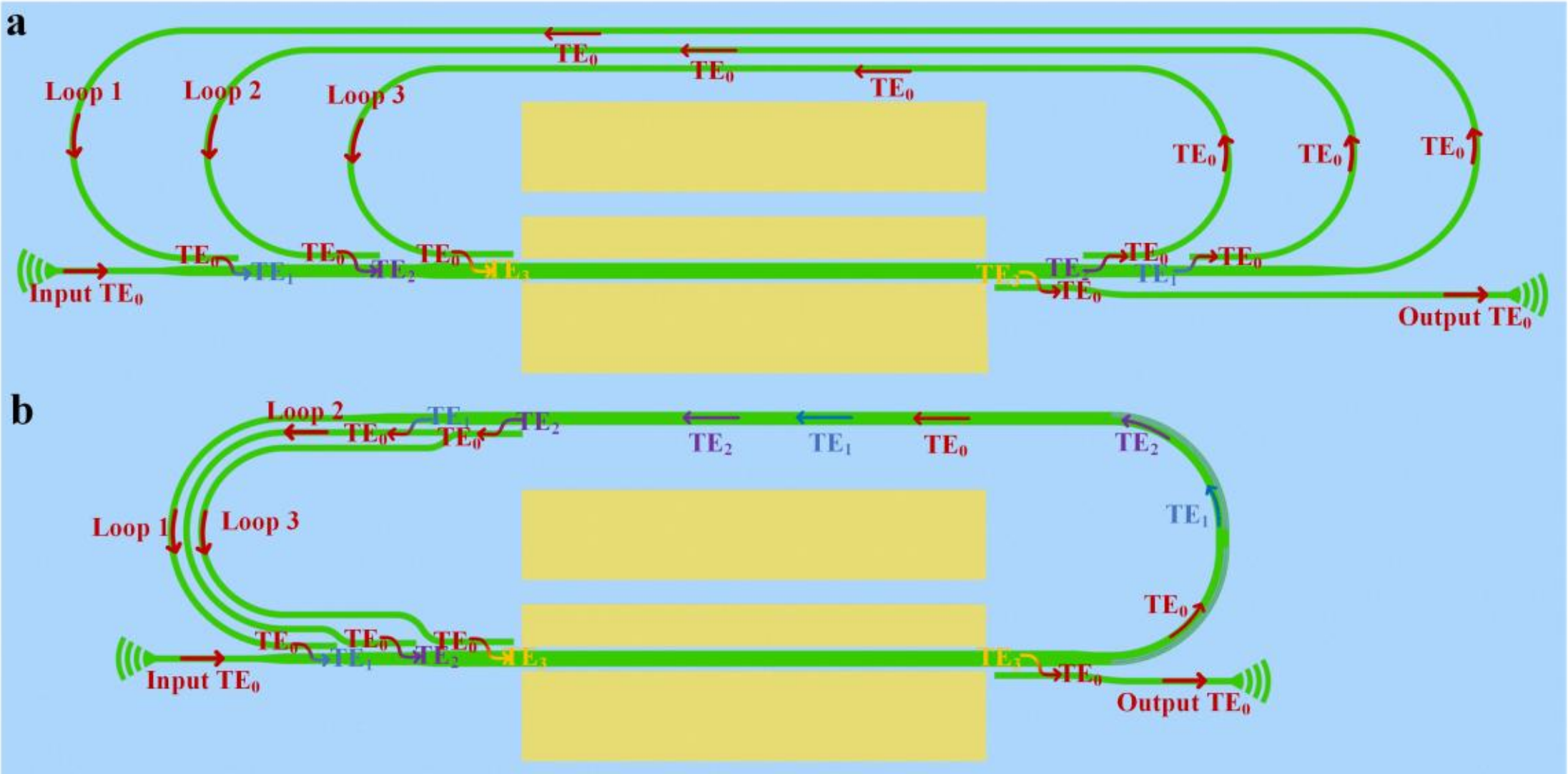

**Figure S2.** Illustrative drawing outlining the structure of **a** the conventional non-resonant recirculating modulator with 3 loops, **b** the novel non-resonant recirculating modulator with 3 loops.

This section elaborates on the detailed operating principles underlying the comparison between the conventional 3-loop recirculating modulator (ORM3) and the proposed 3-loop recirculating modulator (NRM3). For the ORM3, the input $TE_0$ mode light first passes through the modulation region once, then is redirected via a single-mode waveguide for loopback—instead of utilizing multimode waveguide bends (MWBs). Subsequently, the light propagates along a straight single-mode waveguide segment, bends through a circular single-mode waveguide, and is ultimately coupled into the $TE_1$ mode to complete Loop 1. The $TE_1$ mode light undergoes the aforementioned process twice more, eventually being coupled into the $TE_3$ mode at the input side. After its fourth pass through the modulation region, the light is demultiplexed into the $TE_0$ mode for output.

By introducing MWBs, the NRM enhances the time delay difference induced by modal dispersion in waveguides across different loops, while ensuring the physical lengths of the loops remain nearly equal. This approach overcomes the limitation of loop length matching, enabling a more compact device structure and a broader low-$V_\pi$ bandwidth. Specifically, for waveguides with the same width, higher-order modes exhibit larger group refractive indices, resulting in slower light propagation.

Furthermore, loops supporting higher-order modes are positioned closer to the electrodes and thus have relatively shorter physical lengths, whereas loops accommodating lower-order modes are situated farther from the electrodes with correspondingly longer physical lengths. Therefore, despite the use of MWBs to maintain nearly equal physical lengths among loops, the slight physical length differences remaining on the left side of the device can be compensated for by modal dispersion introduced by the multimode waveguides in both the modulation region and the upper path of the device. This compensation ultimately ensures precise equality of all loop delays, maximizing the modulation enhancement factor.

To characterize the constraint of loop length matching on the ORM3, we adopted a loop length ratio of 3:4:5 for Loops 1, 2, and 3 (consistent with previous work[1]) instead of 4:5:6. Simulation results indicate that for a 3 mm-long modulator with the 3:4:5 ratio, the modulation enhancement factor exhibits peak frequencies at 0 GHz and 49.8 GHz. In contrast, the 4:5:6 ratio results in peak frequencies at 0 GHz and 66.4 GHz—where the ideal high-frequency operating point may exceed the 3-dB electro-optic (EO) bandwidth of the modulator. For comparison, the operating frequencies of our 3 mm NRM3 occur at 0, 16.9, 33.8, 50.7, and 67.6 GHz, which significantly alleviates the pressure on the EO bandwidth and demonstrates the advantage of flexible operating frequency points offered by the NRM3. A loop length ratio of 1:2:3 was not selected because, for a 3 mm-long modulator, the modulation enhancement factor would peak at 0 GHz and 16.9 GHz; the latter frequency is excessively low as the maximum operating point, leaving an overly wide frequency margin while introducing unnecessary additional device length.

## Supplementary Note S3: Details in designing the NRM

To achieve precise control over the optical delay in each loop, meticulous design of the total loop length—encompassing modulation arms, MWBs, and mode multiplexers—is imperative. Typically, the total propagation time is calculated using the formula $\Delta T = \Delta L/n_g$, where $\Delta L$ denotes the total length of the waveguide loop and n represents the mean group refractive index of the waveguides. However, since our waveguide loop consists of diverse optical components with varying dimensions and group indices, we perform individual calculations for each device component during the final design to ensure in-phase modulation.

### The novel recirculating modulator structure

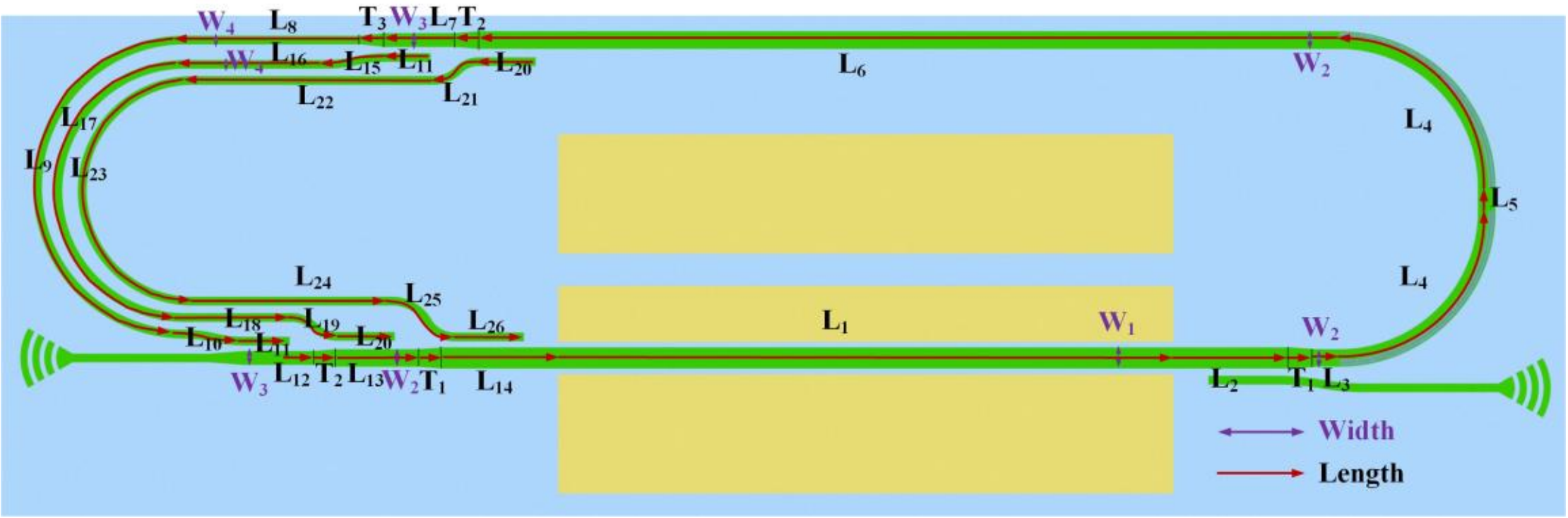

**Figure S3.** The structural parameters of the novel non-resonant recirculating modulator

Here, we illustrate the design principle by focusing on the first loop of the NRM3 as an exemplary case. Within this loop, the optical signal traverses a modulation waveguide $L_1$ and $L_2$ featuring a width of 6.117 μm (operating in the $TE_0$ mode), followed by a taper $L_3$, a 90° slotted multimode waveguide bend (SMWB) $L_4$, looping waveguides of 4.479 μm in width (also in $TE_0$ mode), a straight waveguide $L_5$ with a width of 4.479 μm, another SMWB $L_4$, a straight waveguide $L_6$ with a width of 4.479 μm, a 4.479 μm/2.839 μm taper $T_2$, a straight waveguide $L_7$ with a width of 2.839 μm, a 2.839 μm/1.2 μm taper $T_3$, a straight waveguide $L_8$ with a width of 2.839 μm, a circular waveguide bend $L_9$, a S bend waveguide $L_{10}$, a mode converter region $L_{11}$ and $L_{12}$ and a brief propagation segment in the $TE_1$ mode within 2.839, 4.479 and 6.117 μm waveguides, prior to re-entering the EO modulation region for the subsequent modulation. Additionally, the anisotropic nature of lithium niobate (LN) leads to variations in the group refractive indices of each optical component as the signal travels along diverse crystal orientations. **Supplementary Table 2** outlines the propagation duration for each component, incorporating all aforementioned design aspects. For structures such as tapers, mode multiplexers, and SMWBs, we employ the average group refractive index to ascertain the delay. Notably, the total time delay for each of these three loops is 59.11 ps equivalent to the period of a 16.9 GHz RF signal, thereby guaranteeing precise optical delay and consistent RF phase upon looping back for the second modulation.

Here, we illustrate the design principle using the first loop of NRM3 as an exemplary case. Within this loop, the optical signal propagates through the following components sequentially before re-entering the EO modulation region for subsequent modulation: a modulation waveguide ($L_1/L_2$, width 6.117 µm, $TE_0$ mode), a taper ($L_3$), a 90° slotted multimode waveguide bend (SMWB, $L_4$), looping waveguides (width 4.479 µm, $TE_0$ mode), straight waveguides ($L_5/L_6$, width 4.479 µm), another SMWB ($L_4$), a 4.479 µm/2.839 µm taper ($T_2$), a straight waveguide ($L_7$, width 2.839 µm), a 2.839 µm/1.2 µm taper ($T_3$), a straight waveguide ($L_8$, width 2.839 µm), a circular waveguide bend ($L_9$), an S-bend waveguide ($L_{10}$), mode converter regions ($L_{11}/L_{12}$), and a brief $TE_1$-mode propagation segment in 2.839, 4.479, and 6.117 µm waveguides.

Additionally, the anisotropic nature of LN induces variations in the group refractive index of each optical component when the signal travels along different crystal orientations. **Supplementary Table 2** summarizes the propagation duration of each component, incorporating all the aforementioned design considerations. For structures like tapers, mode multiplexers, and SMWBs, the average group refractive index is used to calculate the delay.

Notably, the total time delay of each of these three loops is 59.11 ps, which equals the period of a 16.9 GHz RF signal. This ensures precise optical delay and consistent RF phase during the second modulation after looping back.

**The Mach-Zehnder modulator**

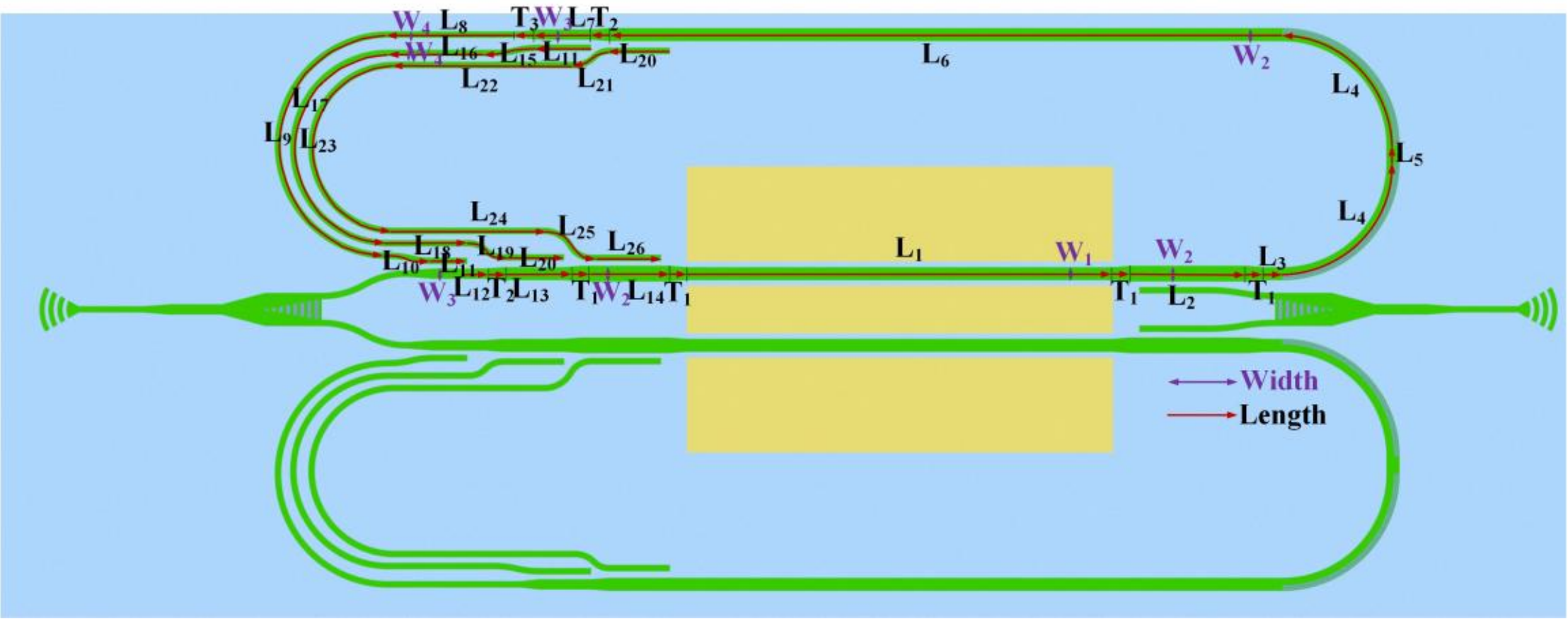

**Figure S4.** The structural parameters of the Mach-Zehnder modulator

Herein, we elucidate the design principle by examining the initial loop of the Mach-Zehnder modulator (MZM) in **Figure S4** as a paradigmatic instance. During the first loop, the optical signal propagates sequentially through the following components before re-entering the EO modulation region for further modulation: a modulation waveguide ($L_1$, width 4.479 μm, $TE_0$ mode), a taper ($T_1$, width transition: 4.479 μm↔6.117 μm↔4.479 μm), a straight waveguide ($L_3$, width 4.479 μm), a 90° slotted multimode waveguide bend (SMWB, $L_4$, width 4.479 μm, $TE_0$ mode), another straight waveguide ($L_5$, width 4.479 μm), a second SMWB, a straight waveguide ($L_6$, width 4.479 μm), a taper ($T_2$, width transition: 4.479 μm→2.839 μm), a straight waveguide ($L_7$, width 2.839 μm), a taper ($T_3$, width transition: 2.84 μm→1.4 μm), a straight waveguide ($L_8$, width 2.84 μm), a circular bend ($L_9$), an S-bend waveguide ($L_{10}$, gradual curvature for loss minimization), a mode converter region ($L_{11}$/$L_{12}$, signal transitions to $TE_1$ mode), and a $TE_1$-mode propagation segment in 2.839 μm, 4.479 μm, and 6.117 μm waveguides.

Furthermore, the anisotropic characteristic of LN introduces variations in the group refractive index of each optical component as the signal propagates through different crystal orientations. **Supplementary Table 3** summarizes the propagation duration of each component, encompassing all aforementioned design features. For structures like tapers, mode multiplexers, and SMWBs, the average group refractive index is used to determine the delay.

Notably, the total time delay of each of these three loops is 44.658 ps, equivalent to the period of a 22.392 GHz RF signal. This ensures precise optical delay and consistent RF phase upon looping back for subsequent modulation.

**The high modulation efficiency phase modulator type II**

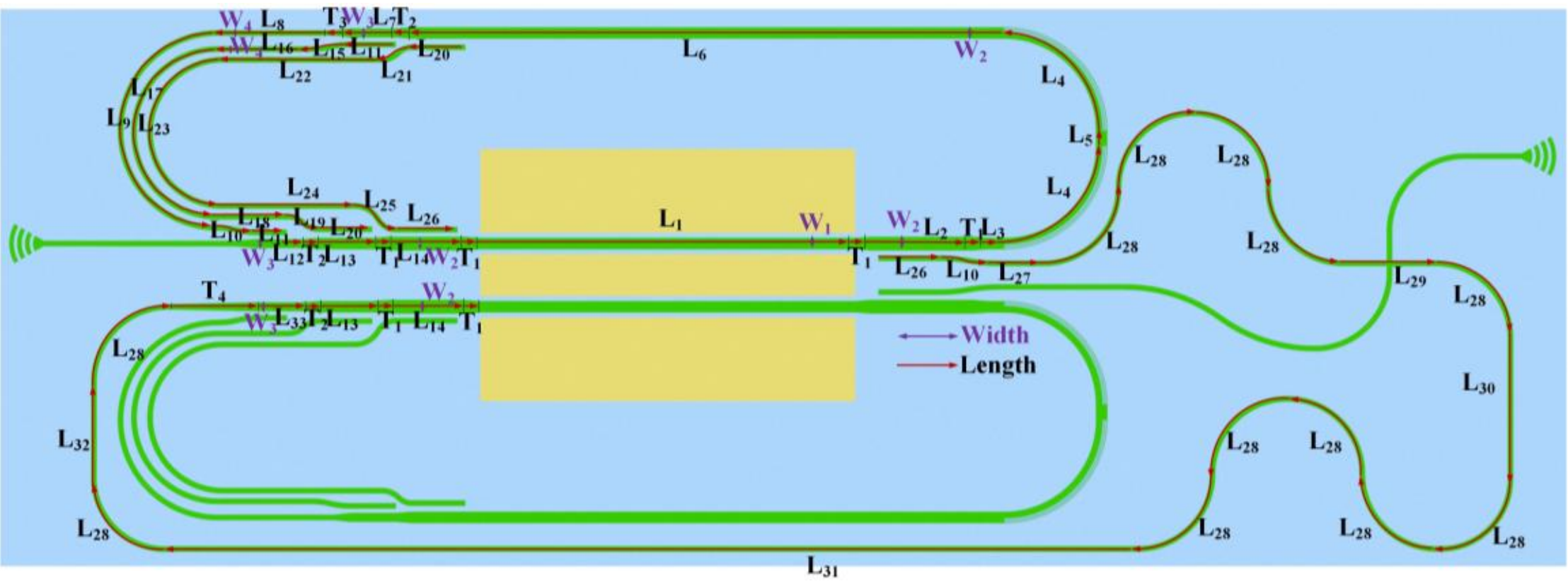

**Figure S5.** The structural parameters of the phase modulator type II

The proposed phase modulator type II (PM2) achieves enhanced modulation efficiency in frequency comb application by simultaneously utilizing both the upper and lower modulation regions of the ground-signal-ground (GSG) electrodes. This dual-region modulation capability significantly improves the device's overall modulation performance, as illustrated in **Figure S5**.

The critical structural distinction between the PM2 resides in the incorporation of a fourth optical loop when light exits the upper modulation region via the $TE_0/TE_3$ mode converter. In this additional loop, the optical signal propagates through an integrated optical delay line prior to entering the lower modulation region of the GSG electrodes. After four successive passes through the lower electrode region—enabling recirculating propagation for enhanced light-matter interaction—the light is reconverted to the $TE_0$ mode by the final $TE_0/TE_3$ mode converter. The modulated signal subsequently traverses a low-loss waveguide crossing and is ultimately coupled off-chip via a grating coupler, completing the phase modulator scheme.

**Supplementary Table 4** compiles the critical design parameters of all structural components. Notably, the fifth loop bridging the upper and lower modulation regions incorporates a waveguide segment that introduces a cumulative time delay of 45.037 ps. In contrast, each of the remaining six loops contributes a delay of 30.024 ps. This precisely engineered differential in path lengths and temporal delays ensures synchronized RF phase alignment and deterministic optical timing for the 33.3 GHz microwave signal during recirculation.

**Supplementary Note S4: Performances of the passive components.**

The structural parameters of the mode converters[2], SMWBs[3], and subwavelength grating (SWG)-assisted Y-branch[4] are illustrated in **Figures S6a–d**. The mode converters share a common gap, with progressively increasing coupling lengths and bus waveguide widths. The SMWBs are configured with two grooves of identical width, while the SWG-assisted Y-branch incorporates a tapered SWG at its junction. Detailed parameters of all aforementioned components are summarized in **Supplementary Table 1.**

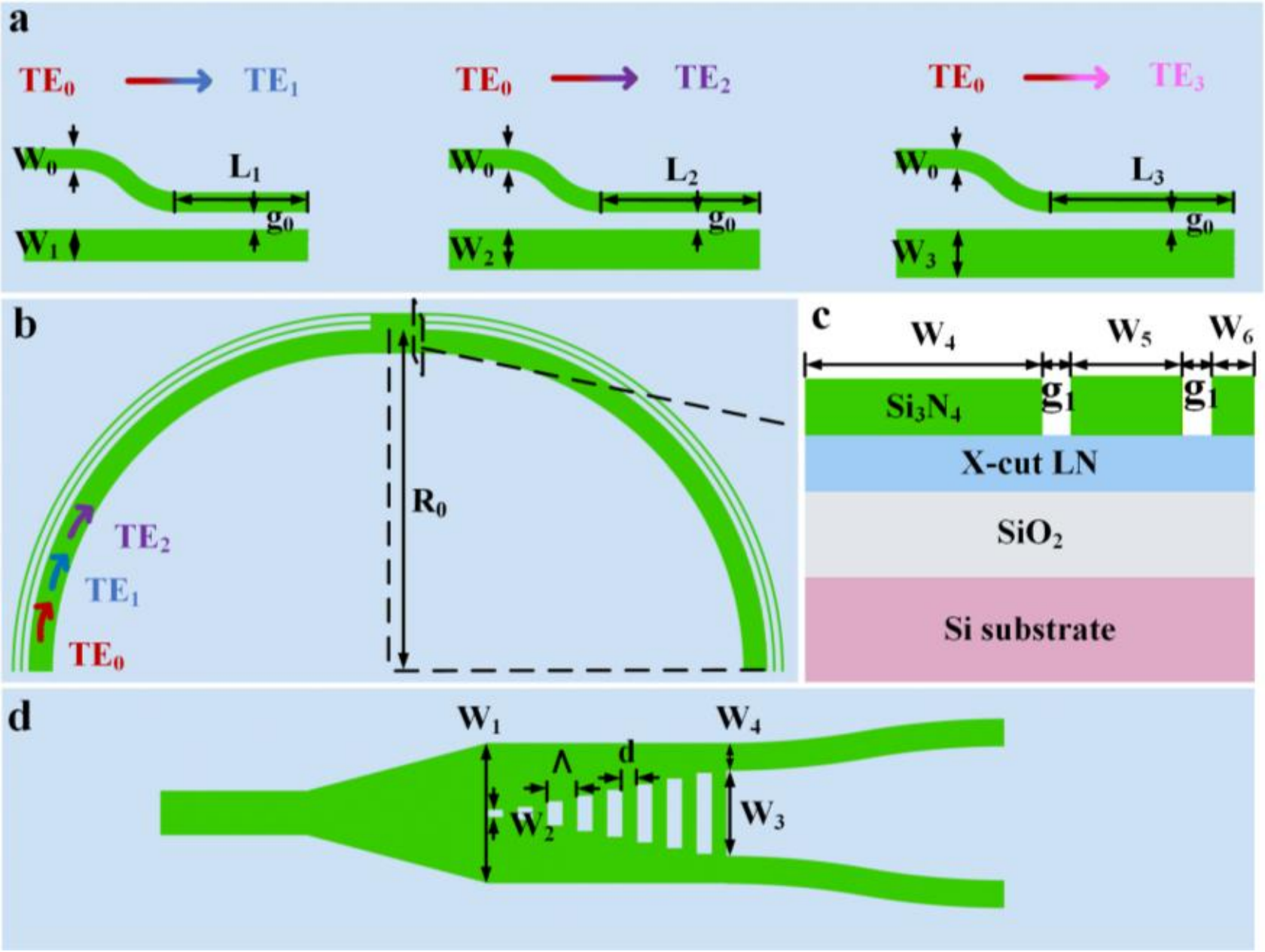

**Figure S6.** The structural parameters of the passive components **a** the structural parameters of the three mode converters, **b, c** the structural parameters of SMWBs **d** the structural parameters of SWG-assisted Y-branch.

The finite difference time domain (FDTD) method simulated the converters' transmission characteristics. Electric field profiles at 1550 nm are shown in **Figures S7a** (i-iii), while simulated transmission results (**Figures S7b** (i-iii)) reveal insertion losses < 0.1 dB, 0.36 dB, and 0.48 dB for $TE_0$/$TE_1$, $TE_0$/$TE_2$, and $TE_0$/$TE_3$ converters, respectively, with inter-modal crosstalk below −18.3 dB, −9.5 dB, and −10 dB across 1500–1600 nm.

Experimental results of the fabricated three-converter device (**Figures S7c** (i-iii)) indicate insertion losses of ~0.13 dB, 0.43 dB, and 0.7 dB for the same converter types, with corresponding inter-mode crosstalk remaining below −18.3 dB, −9.5 dB, and −10 dB within the investigated wavelength range.

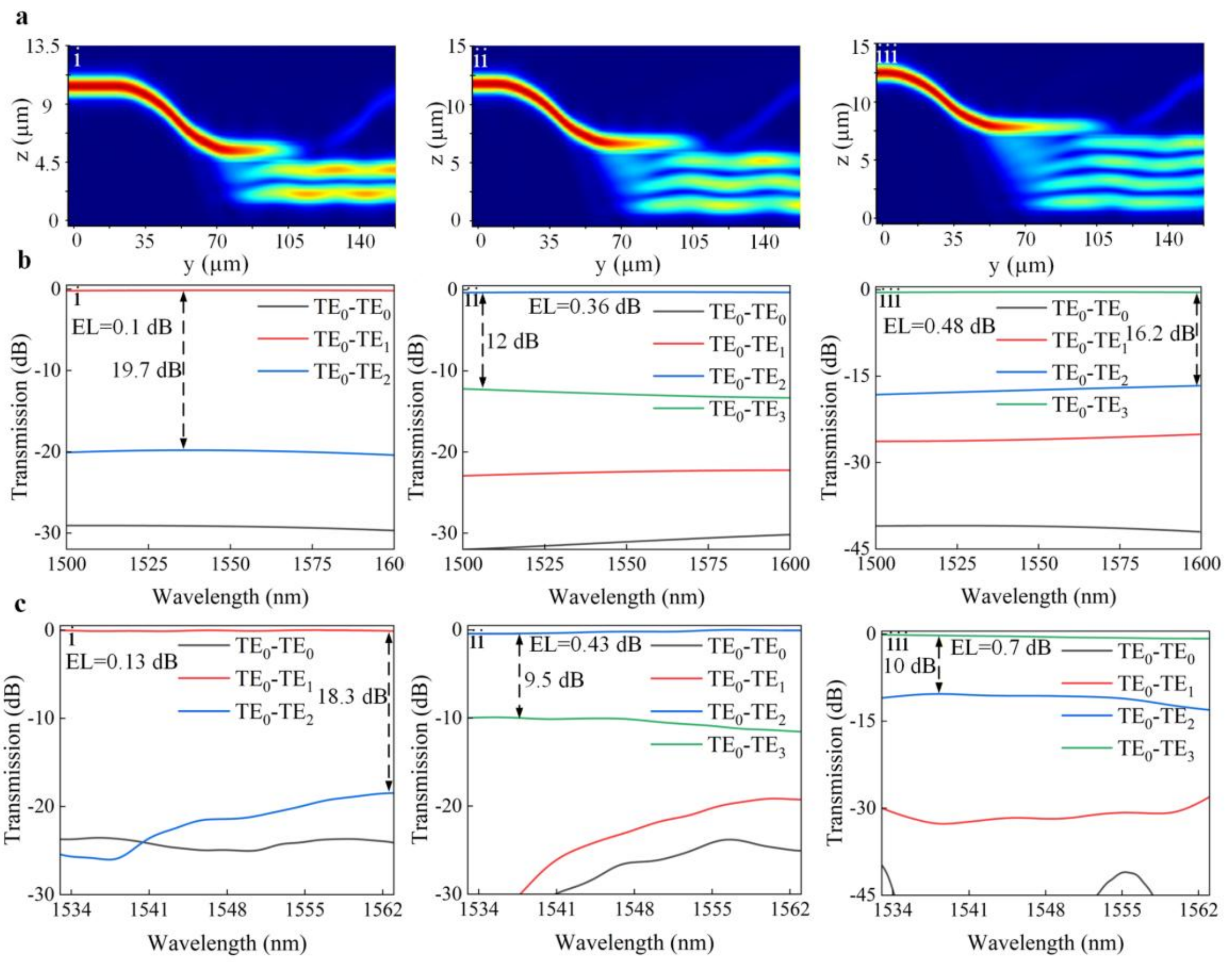

**Figure S7.** The performance of the three mode converters **a** the FDTD simulated results of the **i** $TE_0/TE_1$, **ii** $TE_0/TE_2$, **iii** $TE_0/TE_3$ mode converters, **b** the FDTD calculated transmission results of the **i** $TE_0/TE_1$, **ii** $TE_0/TE_2$, **iii** $TE_0/TE_3$ mode converters, **c** the measured transmission results of the **i** $TE_0/TE_1$, **ii** $TE_0/TE_2$, **iii** $TE_0/TE_3$ mode converters.

The FDTD method was also used to simulate the transmission properties of SMWBs. Electric field distributions of a single SMWB at 1550 nm are depicted in **Figures S8a** (i-iii), and its simulated transmission results (**Figures S8b** (i-iii)) show insertion losses < 0.2 dB, 0.18 dB, and 0.24 dB for $TE_0$, $TE_1$, and $TE_2$ modes, respectively, alongside inter-modal crosstalk below −17.9 dB, −17.3 dB, and −20.8 dB across 1500–1600 nm. Experimental results of the fabricated device with two SMWBs (**Figures S8c** (i-iii)) indicate insertion losses of ~0.43 dB, 0.545 dB, and 0.915 dB for $TE_0$, $TE_1$, and $TE_2$ modes, respectively, with corresponding inter-mode crosstalk below −14.8 dB, −14.3 dB, and −12.1 dB within 1533–1563 nm.

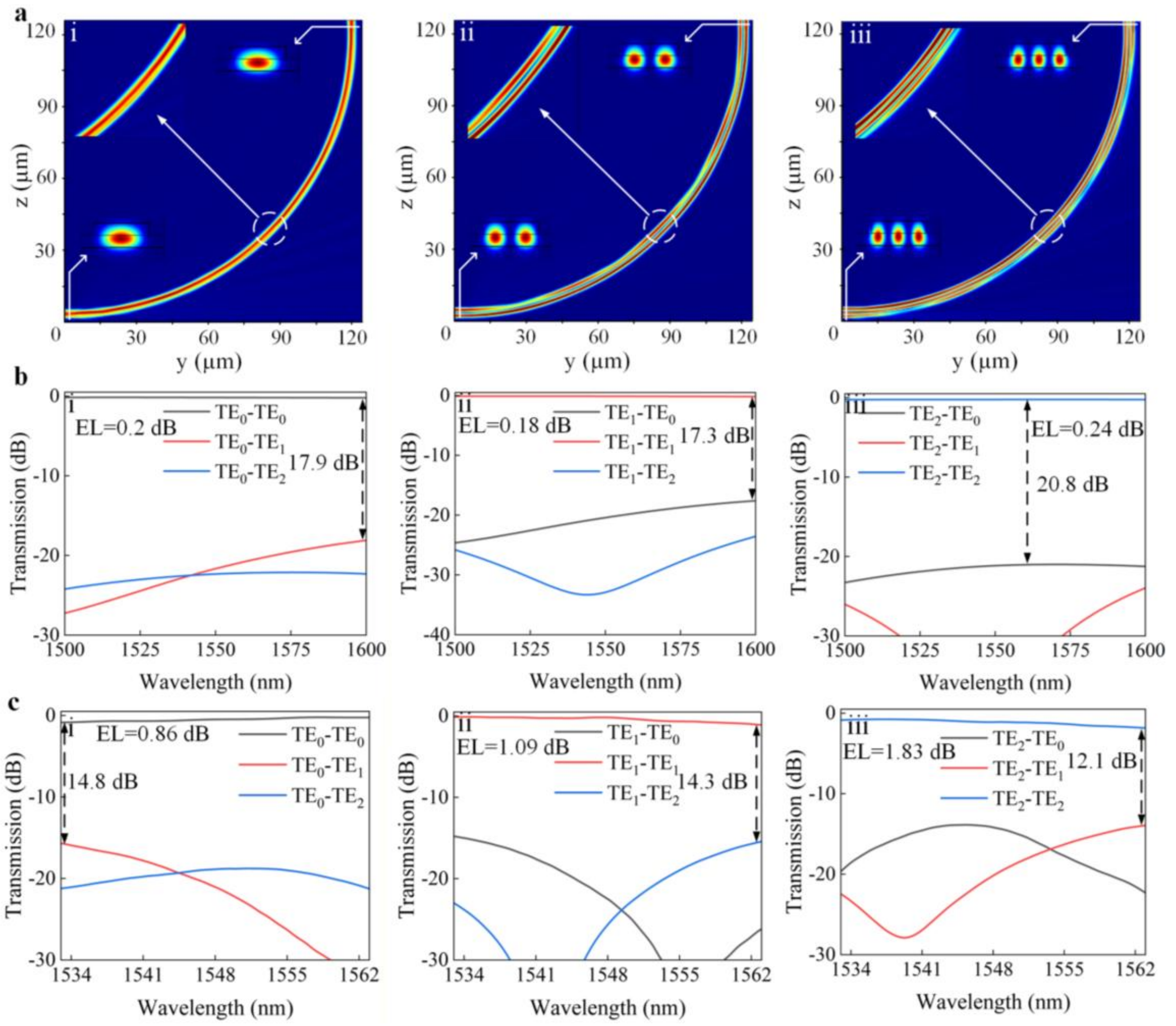

**Figure S8.** The performance of the slotted multimode waveguide bend (SMWB) **a** the FDTD simulated results of the single SMWB **i** $TE_0$ input, **ii** $TE_1$ input, **iii** $TE_2$ input, **b** the FDTD calculated transmission results of the single SMWB **i** $TE_0$ input, **ii** $TE_1$ input, **iii** $TE_2$ input, **c** the measured transmission results of the two SMWBs **i** $TE_0$ input, **ii** $TE_1$ input, **iii** $TE_2$ input.

From the insertion losses of the above devices, it can be derived that our NRM—comprising two $TE_0$/$TE_1$, two $TE_0$/$TE_2$, two $TE_0$/$TE_3$ mode converters, and two 90° SMWBs—introduces a 6.3 dB excess insertion loss compared to a common phase modulator. Utilizing mode converters and multimode waveguide bends with reduced insertion loss will mitigate the excessive insertion loss. For instance, by employing the mode converters and multimode waveguide bends as detailed in previous work[5], we can achieve a significant reduction in the total excess insertion loss, down to as little as 0.454 dB including six mode converters and two 90° MWBs.

**Supplementary Note S5: EO response spectrum simulation**

**The Mach-Zehnder modulator**

Drawing upon the theoretical equation for the modulation bandwidth of the MZM, as cited in previous works[6-8], we deduce the corresponding theoretical equation for the modulation bandwidth of the MZM based on the NRM structure. In our simulations, the constraints on the EO bandwidth stems from the velocity discrepancy between the optical and RF signals and the traveling wave electrode's RF loss. For $TE_0$, $TE_1$, $TE_2$ and $TE_3$ modes, the optical and radio-frequency (RF) signals propagate in unison. Under the application of the RF signal $v(t) = V_{pk}\cos(2\pi f_{RF})$, the phase accumulation associated with the $TE_0$ mode is specified by

$$\begin{cases} \varphi_0(t) = \dfrac{V_{pk}}{V_\pi L}\pi\displaystyle\int_0^L e^{-\frac{\alpha z}{2}}\cos\!\Big(2\pi f_{RF}\big(t + \dfrac{n_{o0}-n_{RF}}{c}z + \dfrac{n_{o3}L}{c}\big)\Big)dz \\ = -\dfrac{V_{pk}}{V_\pi L}\pi\sqrt{\dfrac{e^{-\alpha L} + 1 - 2e^{-\frac{\alpha L}{2}}\cos\big(\frac{2\pi f_{RF}(n_{RF}-n_{o0})L}{c}\big)}{\big(\frac{2\pi f_{RF}(n_{RF}-n_{o0})}{c}\big)^2 + \big(\frac{\alpha}{2}\big)^2}}\cos\!\Big(2\pi f_{RF}\big(t + \dfrac{n_{o3}L}{c}\big) + \phi_0\Big) \\ \phi_0 = \arctan\left(\dfrac{e^{-\frac{\alpha L}{2}}\alpha\cos\big(\frac{2\pi f_{RF}(n_{RF}-n_{o0})L}{c}\big) - \dfrac{4e^{-\frac{\alpha L}{2}}\pi f_{RF}(n_{RF}-n_{o0})\sin\big(\frac{2\pi f_{RF}(n_{RF}-n_{o0})L}{c}\big)}{c} - \alpha}{e^{-\frac{\alpha L}{2}}\alpha\sin\big(\frac{2\pi f_{RF}(n_{RF}-n_{o0})L}{c}\big) + \dfrac{4e^{-\frac{\alpha L}{2}}\pi f_{RF}(n_{RF}-n_{o0})\cos\big(\frac{2\pi f_{RF}(n_{RF}-n_{o0})L}{c}\big)}{c} - \dfrac{4\pi f_{RF}(n_{RF}-n_{o0})}{c}}\right) \end{cases}$$

(6)

where $V_{pk}$ is the peak voltage, $f_{RF}$ is the microwave frequency, $V_\pi$ is the half-wave voltage, $L = 2\ mm$ is the waveguide length of the modulation region, $c$ is the speed of light in free space, $n_{o0} = 2.1935$ is the group index of $TE_0$ mode, $n_{o3} = 2.2838$ is the group index of $TE_3$ mode, $n_{RF}$ is the group index of the RF signal ($n_{RF} = 2.077$ at 40 GHz), $\phi_0$ is the additional phase shift caused by microwave loss of $TE_0$ mode. $\alpha$ is the microwave power attenuation constant, which is shown in **Fig S9**.

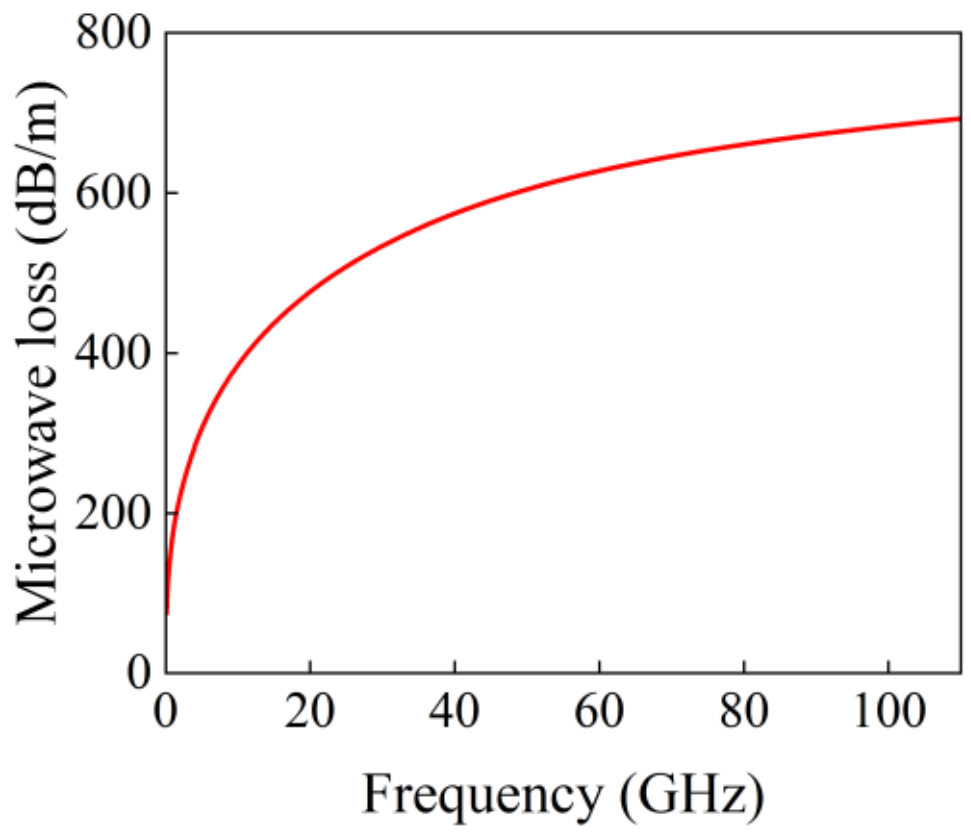

**Figure S9.** The simulated microwave loss from 10 MHz to 110 GHz.

The phase accumulation associated with the $TE_1$ mode is specified by

$$
\begin{cases}
\varphi_1(t) = \dfrac{V_{pk}}{V_\pi L}\pi\displaystyle\int_0^L e^{-\frac{\alpha z}{2}}\cos\left(2\pi f_{RF}\left(t+\dfrac{n_{o1}-n_{RF}}{c}z+\Delta T_{loop}+\dfrac{n_{o3}L}{c}\right)\right)dz \\
= -\dfrac{V_{pk}}{V_\pi L}\pi\sqrt{\dfrac{e^{-\alpha L}+1-2e^{-\frac{\alpha L}{2}}\cos\left(\dfrac{2\pi f_{RF}(n_{RF}-n_{o1})L}{c}\right)}{\left(\dfrac{2\pi f_{RF}(n_{RF}-n_{o1})}{c}\right)^2+\left(\dfrac{\alpha}{2}\right)^2}}\cos\left(2\pi f_{RF}\left(t+\Delta T_{loop}+\dfrac{n_{o3}L}{c}\right)+\phi_1\right) \\
\phi_1 = \arctan\left(\dfrac{e^{-\frac{\alpha L}{2}}\alpha\cos\left(\dfrac{2\pi f_{RF}(n_{RF}-n_{o1})L}{c}\right)-\dfrac{4e^{-\frac{\alpha L}{2}}\pi f_{RF}(n_{RF}-n_{o1})\sin\left(\dfrac{2\pi f_{RF}(n_{RF}-n_{o1})L}{c}\right)}{c}-\alpha}{e^{-\frac{\alpha L}{2}}\alpha\sin\left(\dfrac{2\pi f_{RF}(n_{RF}-n_{o1})L}{c}\right)+\dfrac{4e^{-\frac{\alpha L}{2}}\pi f_{RF}(n_{RF}-n_{o1})\cos\left(\dfrac{2\pi f_{RF}(n_{RF}-n_{o1})L}{c}\right)}{c}-\dfrac{4\pi f_{RF}(n_{RF}-n_{o1})}{c}}\right)
\end{cases}
$$

(7)

where $\Delta T_{loop} = 44.658\ ps$ is the total delay time of the loop, $n_{o1} = 2.2133$ is the group index of $TE_1$ mode, $\phi_1$ is the additional phase shift caused by microwave loss of $TE_1$ mode.

The phase accumulation associated with the $TE_2$ mode is specified by

$$\begin{cases} \varphi_2(t) = \frac{V_{pk}}{V_\pi L}\pi\int_0^L e^{-\frac{\alpha z}{2}}\cos(2\pi f_{RF}(t+\frac{n_{o2}-n_{RF}}{c}z+2\Delta T_{loop}+\frac{n_{o3}L}{c}))dz \\ = -\frac{V_{pk}}{V_\pi L}\pi\sqrt{\frac{e^{-\alpha L}+1-2e^{-\frac{\alpha L}{2}}\cos(\frac{2\pi f_{RF}(n_{RF}-n_{o2})L}{c})}{(\frac{2\pi f_{RF}(n_{RF}-n_{o2})}{c})^2+(\frac{\alpha}{2})^2}}\cos(2\pi f_{RF}(t+2\Delta T_{loop}+\frac{n_{o3}L}{c})+\phi_2) \\ \phi_2 = \arctan(\frac{e^{-\frac{\alpha L}{2}}\alpha\cos(\frac{2\pi f_{RF}(n_{RF}-n_{o2})L}{c})-\frac{4e^{-\frac{\alpha L}{2}}\pi f_{RF}(n_{RF}-n_{o2})\sin(\frac{2\pi f_{RF}(n_{RF}-n_{o2})L}{c})}{c}-\alpha}{e^{-\frac{\alpha L}{2}}\alpha\sin(\frac{2\pi f_{RF}(n_{RF}-n_{o2})L}{c})+\frac{4e^{-\frac{\alpha L}{2}}\pi f_{RF}(n_{RF}-n_{o2})\cos(\frac{2\pi f_{RF}(n_{RF}-n_{o2})L}{c})}{c}-\frac{4\pi f_{RF}(n_{RF}-n_{o2})}{c}}) \end{cases}$$

(8)

where $n_{o2}=2.2452$ is the group index of TE2 mode. $\phi_2$ is the additional phase shift caused by microwave loss of TE2 mode.

The phase accumulation associated with the TE3 mode is specified by

$$\begin{cases} \varphi_3(t) = \frac{V_{pk}}{V_\pi L}\pi\int_0^L e^{-\frac{\alpha z}{2}}\cos(2\pi f_{RF}(t+\frac{n_{o3}-n_{RF}}{c}z+3\Delta T_{loop}+\frac{n_{o3}L}{c}))dz \\ = -\frac{V_{pk}}{V_\pi L}\pi\sqrt{\frac{e^{-\alpha L}+1-2e^{-\frac{\alpha L}{2}}\cos(\frac{2\pi f_{RF}(n_{RF}-n_{o3})L}{c})}{(\frac{2\pi f_{RF}(n_{RF}-n_{o3})}{c})^2+(\frac{\alpha}{2})^2}}\cos(2\pi f_{RF}(t+3\Delta T_{loop}+\frac{n_{o3}L}{c})+\phi_3) \\ \phi_3 = \arctan(\frac{e^{-\frac{\alpha L}{2}}\alpha\cos(\frac{2\pi f_{RF}(n_{RF}-n_{o3})L}{c})-\frac{4e^{-\frac{\alpha L}{2}}\pi f_{RF}(n_{RF}-n_{o3})\sin(\frac{2\pi f_{RF}(n_{RF}-n_{o3})L}{c})}{c}-\alpha}{e^{-\frac{\alpha L}{2}}\alpha\sin(\frac{2\pi f_{RF}(n_{RF}-n_{o3})L}{c})+\frac{4e^{-\frac{\alpha L}{2}}\pi f_{RF}(n_{RF}-n_{o3})\cos(\frac{2\pi f_{RF}(n_{RF}-n_{o3})L}{c})}{c}-\frac{4\pi f_{RF}(n_{RF}-n_{o3})}{c}}) \end{cases}$$

(9)

where $n_{o3}=2.2838$ is the group index of TE3 mode, $\phi_3$ is the additional phase shift caused by microwave loss of TE3 mode.

Consequently, we obtained the overall phase displacement φ(t) as our result

$$
\begin{cases}
\varphi(t)=\varphi_0(t)+\varphi_1(t)+\varphi_2(t)+\varphi_3(t) \\
=-\dfrac{V_{pk}}{V_\pi L}\pi(A\sin(2\pi f_{RF}t)+B\sin(2\pi f_{RF}t+\theta_1)+C\sin(2\pi f_{RF}t+\theta_2)+D\sin(2\pi f_{RF}t+\theta_3)) \\
=-\dfrac{V_{pk}}{V_\pi L}\pi\sqrt{(A+B\cos(\theta_1)+C\cos(\theta_2)+D\cos(\theta_3))^2+(B\sin(\theta_1)+C\sin(\theta_2)+D\sin(\theta_3))^2} \\
\quad \sin(2\pi f_{RF}t+\theta_4) \\
=-\dfrac{V_{pk}}{V_\pi L}\pi E(f_{RF})\sin(2\pi f_{RF}t+\theta_4) \\
A=\sqrt{\dfrac{e^{-\alpha L}+1-2e^{-\frac{\alpha L}{2}}\cos(\frac{2\pi f_{RF}(n_{RF}-n_{o0})L}{c})}{(\frac{2\pi f_{RF}(n_{RF}-n_{o0})}{c})^2+(\frac{\alpha}{2})^2}} \\
B=\sqrt{\dfrac{e^{-\alpha L}+1-2e^{-\frac{\alpha L}{2}}\cos(\frac{2\pi f_{RF}(n_{RF}-n_{o1})L}{c})}{(\frac{2\pi f_{RF}(n_{RF}-n_{o1})}{c})^2+(\frac{\alpha}{2})^2}} \\
C=\sqrt{\dfrac{e^{-\alpha L}+1-2e^{-\frac{\alpha L}{2}}\cos(\frac{2\pi f_{RF}(n_{RF}-n_{o2})L}{c})}{(\frac{2\pi f_{RF}(n_{RF}-n_{o2})}{c})^2+(\frac{\alpha}{2})^2}} \\
D=\sqrt{\dfrac{e^{-\alpha L}+1-2e^{-\frac{\alpha L}{2}}\cos(\frac{2\pi f_{RF}(n_{RF}-n_{o3})L}{c})}{(\frac{2\pi f_{RF}(n_{RF}-n_{o3})}{c})^2+(\frac{\alpha}{2})^2}} \\
\theta_1=2\pi f_{RF}\Delta\mathrm{T}_{loop}+\phi_1-\phi_0 \\
\theta_2=4\pi f_{RF}\Delta\mathrm{T}_{loop}+\phi_2-\phi_0 \\
\theta_3=6\pi f_{RF}\Delta\mathrm{T}_{loop}+\phi_3-\phi_0 \\
\theta_4=\arctan(\dfrac{B\sin(\theta_1)+C\sin(\theta_2)+D\sin(\theta_3)}{A+B\cos(\theta_1)+C\cos(\theta_2)+D\cos(\theta_3)}) \\
E(f_{RF})=\sqrt{(A+B\cos(\theta_1)+C\cos(\theta_2)+D\cos(\theta_3))^2+(B\sin(\theta_1)+C\sin(\theta_2)+D\sin(\theta_3))^2}
\end{cases}
\tag{10}
$$

Subsequently, the EO response spectrum $S_{EO}(f_{RF})$

$$
S_{EO}(f_{RF})=10\cdot log_{10}\left|\frac{J_1(-\frac{V_{pk}}{V_\pi L}\pi E(f_{RF}))}{J_1(-\frac{V_{pk}}{V_\pi L}\pi E(0))}\right|^2 \tag{11}
$$

where $J_1(\cdot)$ is the first-order Bessel function of the first kind.

By substituting the parameters into Equation (11), the calculated EO bandwidth of the proposed MZM is shown in **Fig. S10** with a red dotted curve. The measured EO response is described with a blue solid line.

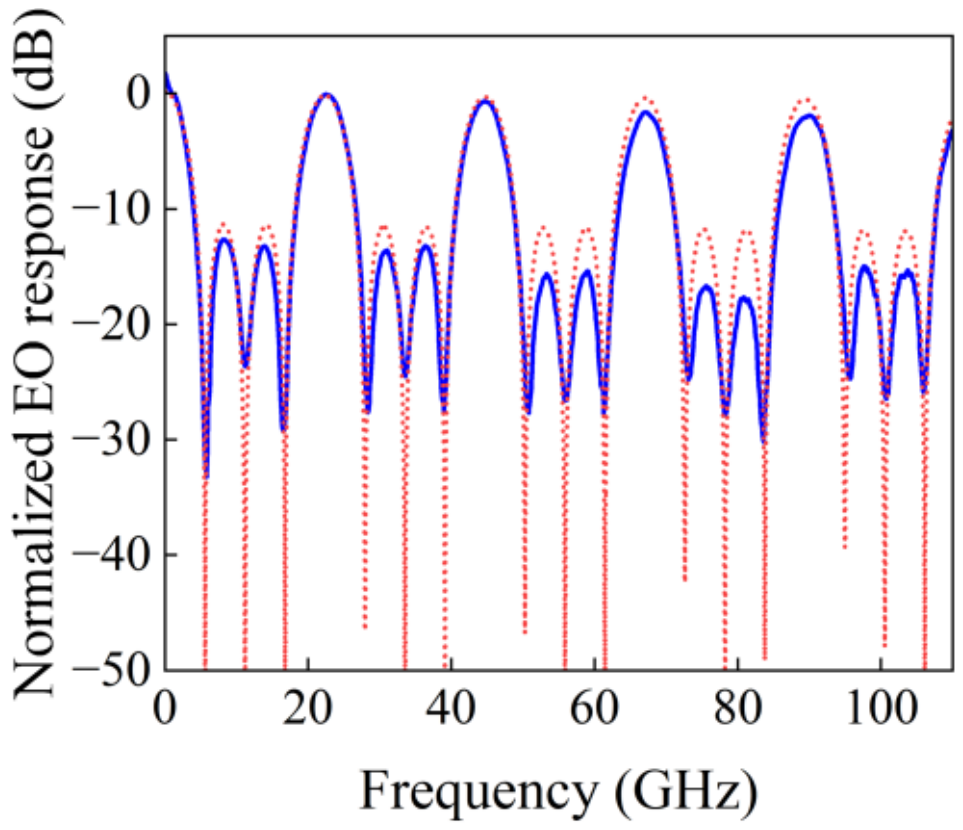

**Figure S10.** Measured (blue solid line) and simulated (red dotted curve) EO bandwidth from 10 MHz to 110 GHz.

It can be seen that the simulation results are generally in accordance with the experimental data with the exception of the absence of small sidelobes next to the main resonant peak. We believe that it is suppressed by the amplitude and phase responses of spectral fluctuations resulting from the crosstalk between the mode converters and the SMWBs. The excess decay of the measured EO bandwidth in high frequency range may arise from the impedance mismatching or other factors. We also simulated the loop delay time shift caused by the microwave loss using the equation

$$\begin{cases} \Delta T_1 = \dfrac{(\phi_2 - \phi_1)}{2\pi f_{RF}} \\ \Delta T_2 = \dfrac{(\phi_3 - \phi_1)}{2\pi f_{RF}} \\ \Delta T_3 = \dfrac{(\phi_4 - \phi_1)}{2\pi f_{RF}} \end{cases} \tag{12}$$

The result is about $\Delta T_1 = 0.01046$ ps, $\Delta T_2 = 0.02732$ ps, $\Delta T_3 = 0.04772$ ps from 0 to 110 GHz. The loop delay time shift caused by microwave loss is very small, which is negligible compared to 44.658 ps. Therefore, we can neglect the loop delay time shift caused by the microwave loss.

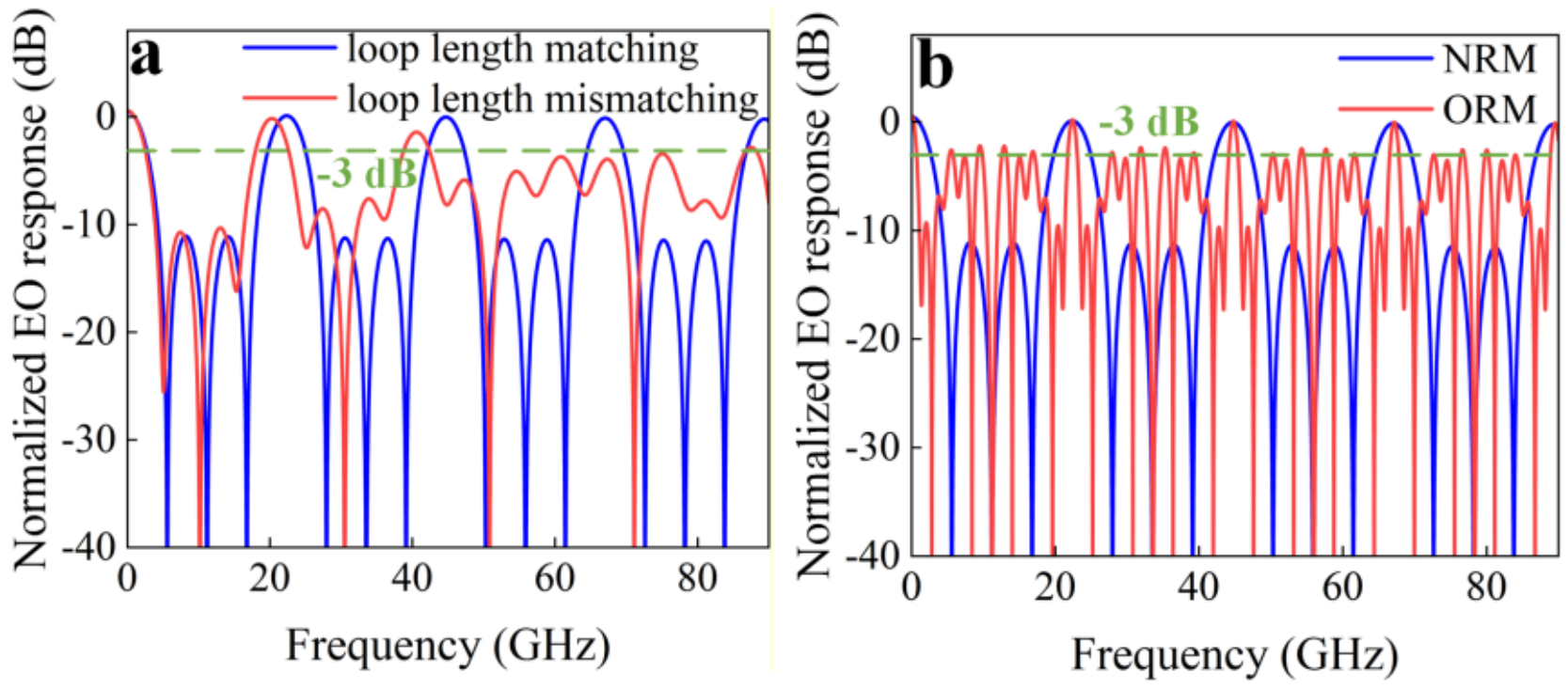

**Figure S11.** Simulated EO bandwidth from 0 to 90 GHz **a** the loop length matching condition (blue line) and the loop length mismatching condition (red line), **b** the novel recirculating modulator (blue line) and the old recirculating modulator (red line). ORM3: old recirculating modulator with three loops, NRM3: novel recirculating modulator with three loops.

To ascertain the significance of loop length matching , we undertake simulations using equation (11), as shown in **Figure S11a**, comparing a well-matched RM configured with three loops maintaining a loop delay time ratio of 1:1:1, against a mismatched RM comprising three loops with a loop delay time ratio of 1:1.1:1.2. $\Delta T_{\mathrm{loop}}$ of the ORM and the NRM is set to 44.658 ps to ensure that the target working resonant frequency is precisely 22.392 GHz. We can observe a notable shift in the resonant frequency of the mismatched RM, with a clear and distinct reduction towards a lower frequency. Furthermore, the resonant peak undergoes a dramatic decrease in magnitude, indicating a significant deterioration in the resonance characteristics of the mismatched RM compared to its well-matched counterpart.

In order to showcase the advantages of our NRM compared to ORM in enhancing the low-$V_\pi$ bandwidth, we conducted simulations using equation (11), as shown in **Figure S11b**, comparing a configuration of NRM comprising three loops with a loop delay time ratio of 1:1:1, against an ORM configuration consisting of three loops with a loop delay time ratio of 3:4:5. $\Delta T_{loop}$ of the ORM and the NRM is set to 44.658 ps to ensure that the target working resonant frequency is precisely 22.392 GHz. We can observe that the resonant peaks of the NRM and the ORM are identical. However, the frequency range above the -3 dB line for the NRM is significantly broader than that of the ORM, spanning approximately 5.08 GHz compared to just 1.26 GHz. For both the NRM and ORM EO responses, the small sidelobes between the two resonant frequencies will be suppressed by the amplitude and phase responses of spectral fluctuations resulting from the crosstalk.

**The phase modulator type II**

Based on the theoretical framework established in Equations (6)-(9), we propose a comprehensive model to analyze the EO response characteristics of the PM2. The device architecture employs a cascaded configuration where the optical signal undergoes sequential modulation through two distinct active regions. Specifically, the light initially propagates through the upper modulation region before being optically coupled into the lower modulation region via an optimized waveguide transition. In the second stage, the recoupled optical signal modulated by the lower electrode structure while maintaining the $TE_0$ mode, thereby enabling additional phase modulation through the secondary EO interaction. The phase accumulation associated with the $TE_0$ mode in the lower modulation region is specified by:

$$\begin{cases} \varphi_0'(t) = \dfrac{V_{pk}}{V_\pi L}\pi\displaystyle\int_0^L e^{-\frac{\alpha z}{2}}\cos(\pi + 2\pi f_{RF}(t + \dfrac{n_{o0} - n_{RF}}{c}z + 3\Delta\mathrm{T}_{loop} + \Delta\mathrm{T}_{cross} + \dfrac{n_{o3}L}{c}))dz \\ = -\dfrac{V_{pk}}{V_\pi L}\pi\sqrt{\dfrac{e^{-\alpha L} + 1 - 2e^{-\frac{\alpha L}{2}}\cos(\frac{2\pi f_{RF}(n_{RF} - n_{o0})L}{c})}{(\frac{2\pi f_{RF}(n_{RF} - n_{o0})}{c})^2 + (\frac{\alpha}{2})^2}}\cos(\pi + 2\pi f_{RF}(t + 3\Delta\mathrm{T}_{loop} + \Delta\mathrm{T}_{cross} - \dfrac{n_{o3}L}{c}) + \phi_0) \\ \phi_0 = \arctan\left(\dfrac{e^{-\frac{\alpha L}{2}}\alpha\cos(\frac{2\pi f_{RF}(n_{RF} - n_{o0})L}{c}) - \dfrac{4e^{-\frac{\alpha L}{2}}\pi f_{RF}(n_{RF} - n_{o0})\sin(\frac{2\pi f_{RF}(n_{RF} - n_{o0})L}{c})}{c} - \alpha}{e^{-\frac{\alpha L}{2}}\alpha\sin(\frac{2\pi f_{RF}(n_{RF} - n_{o0})L}{c}) + \dfrac{4e^{-\frac{\alpha L}{2}}\pi f_{RF}(n_{RF} - n_{o0})\cos(\frac{2\pi f_{RF}(n_{RF} - n_{o0})L}{c})}{c} - \dfrac{4\pi f_{RF}(n_{RF} - n_{o0})}{c}}\right) \end{cases}$$

(13)

where $\Delta\mathrm{T}_{cross} = 450.371\,ps$ is total delay time of the loop bridging the upper and lower modulation regions, $\Delta\mathrm{T}_{loop} = 30.025\,ps$ is total delay time of the each of the remaining six loops, $L = 1\,mm$ is the electrode length of the high modulation efficiency phase modulator.

Similarly, the $TE_1$, $TE_2$ and $TE_3$ mode's phase accumulation within the lower modulation domain is quantitatively defined through the following expression:

$$\begin{cases}
\varphi_1'(t) = \dfrac{V_{pk}}{V_\pi L}\pi\displaystyle\int_0^L e^{-\frac{\alpha z}{2}}\cos(\pi + 2\pi f_{RF}(t + \frac{n_{o1}-n_{RF}}{c}z + 4\Delta T_{loop} + \Delta T_{cross} + \frac{n_{o3}L}{c}))dz \\
= -\dfrac{V_{pk}}{V_\pi L}\pi\sqrt{\dfrac{e^{-\alpha L} + 1 - 2e^{-\frac{\alpha L}{2}}\cos(\frac{2\pi f_{RF}(n_{RF}-n_{o1})L}{c})}{(\frac{2\pi f_{RF}(n_{RF}-n_{o1})}{c})^2 + (\frac{\alpha}{2})^2}}\cos(\pi + 2\pi f_{RF}(t + 4\Delta T_{loop} + \Delta T_{cross} + \frac{n_{o3}L}{c}) + \phi_1) \\
\phi_1 = \arctan\left(\dfrac{e^{-\frac{\alpha L}{2}}\alpha\cos(\frac{2\pi f_{RF}(n_{RF}-n_{o1})L}{c}) - \frac{4e^{-\frac{\alpha L}{2}}\pi f_{RF}(n_{RF}-n_{o1})\sin(\frac{2\pi f_{RF}(n_{RF}-n_{o1})L}{c})}{c} - \alpha}{e^{-\frac{\alpha L}{2}}\alpha\sin(\frac{2\pi f_{RF}(n_{RF}-n_{o1})L}{c}) + \frac{4e^{-\frac{\alpha L}{2}}\pi f_{RF}(n_{RF}-n_{o1})\cos(\frac{2\pi f_{RF}(n_{RF}-n_{o1})L}{c})}{c} - \frac{4\pi f_{RF}(n_{RF}-n_{o1})}{c}}\right) \\
\varphi_2'(t) = \dfrac{V_{pk}}{V_\pi L}\pi\displaystyle\int_0^L e^{-\frac{\alpha z}{2}}\cos(\pi + 2\pi f_{RF}(t + \frac{n_{o2}-n_{RF}}{c}z + 5\Delta T_{loop} + \Delta T_{cross} + \frac{n_{o3}L}{c}))dz \\
= -\dfrac{V_{pk}}{V_\pi L}\pi\sqrt{\dfrac{e^{-\alpha L} + 1 - 2e^{-\frac{\alpha L}{2}}\cos(\frac{2\pi f_{RF}(n_{RF}-n_{o2})L}{c})}{(\frac{2\pi f_{RF}(n_{RF}-n_{o2})}{c})^2 + (\frac{\alpha}{2})^2}}\cos(\pi + 2\pi f_{RF}(t + 5\Delta T_{loop} + \Delta T_{cross} + \frac{n_{o3}L}{c}) + \phi_2) \\
\phi_2 = \arctan\left(\dfrac{e^{-\frac{\alpha L}{2}}\alpha\cos(\frac{2\pi f_{RF}(n_{RF}-n_{o2})L}{c}) - \frac{4e^{-\frac{\alpha L}{2}}\pi f_{RF}(n_{RF}-n_{o2})\sin(\frac{2\pi f_{RF}(n_{RF}-n_{o2})L}{c})}{c} - \alpha}{e^{-\frac{\alpha L}{2}}\alpha\sin(\frac{2\pi f_{RF}(n_{RF}-n_{o2})L}{c}) + \frac{4e^{-\frac{\alpha L}{2}}\pi f_{RF}(n_{RF}-n_{o2})\cos(\frac{2\pi f_{RF}(n_{RF}-n_{o2})L}{c})}{c} - \frac{4\pi f_{RF}(n_{RF}-n_{o2})}{c}}\right) \\
\varphi_3'(t) = \dfrac{V_{pk}}{V_\pi L}\pi\displaystyle\int_0^L e^{-\frac{\alpha z}{2}}\cos(\pi + 2\pi f_{RF}(t + \frac{n_{o3}-n_{RF}}{c}z + 6\Delta T_{loop} + \Delta T_{cross} + \frac{n_{o3}L}{c}))dz \\
= -\dfrac{V_{pk}}{V_\pi L}\pi\sqrt{\dfrac{e^{-\alpha L} + 1 - 2e^{-\frac{\alpha L}{2}}\cos(\frac{2\pi f_{RF}(n_{RF}-n_{o3})L}{c})}{(\frac{2\pi f_{RF}(n_{RF}-n_{o3})}{c})^2 + (\frac{\alpha}{2})^2}}\cos(\pi + 2\pi f_{RF}(t + 6\Delta T_{loop} + \Delta T_{cross} + \frac{n_{o3}L}{c}) + \phi_3) \\
\phi_3 = \arctan\left(\dfrac{e^{-\frac{\alpha L}{2}}\alpha\cos(\frac{2\pi f_{RF}(n_{RF}-n_{o3})L}{c}) - \frac{4e^{-\frac{\alpha L}{2}}\pi f_{RF}(n_{RF}-n_{o3})\sin(\frac{2\pi f_{RF}(n_{RF}-n_{o3})L}{c})}{c} - \alpha}{e^{-\frac{\alpha L}{2}}\alpha\sin(\frac{2\pi f_{RF}(n_{RF}-n_{o3})L}{c}) + \frac{4e^{-\frac{\alpha L}{2}}\pi f_{RF}(n_{RF}-n_{o3})\cos(\frac{2\pi f_{RF}(n_{RF}-n_{o3})L}{c})}{c} - \frac{4\pi f_{RF}(n_{RF}-n_{o3})}{c}}\right)
\end{cases}$$

(14)

Based on the theoretical framework established in Equations (6)(7)(8)(9)(13)(14), we can derive the overall phase displacement φ(t) as our result:

$$
\begin{cases}
\varphi(t) = \varphi_0(t) + \varphi_1(t) + \varphi_2(t) + \varphi_3(t) + \varphi_0'(t) + \varphi_1'(t) + \varphi_2'(t) + \varphi_3'(t) \\
= -\frac{V_{pk}}{V_\pi L}\pi(A\sin(2\pi f_{RF}t) + B\sin(2\pi f_{RF}t + \theta_1) + C\sin(2\pi f_{RF}t + \theta_2) + D\sin(2\pi f_{RF}t + \theta_3) + \\
+B\sin(2\pi f_{RF}t + \theta_1') + C\sin(2\pi f_{RF}t + \theta_2') + D\sin(2\pi f_{RF}t + \theta_3')) \\
= -\frac{V_{pk}}{V_\pi L}\pi\sqrt{(A + B\cos(\theta_1) + C\cos(\theta_2) + D\cos(\theta_3) + A\cos(\theta_{cross}) + B\cos(\theta_1') + C\cos(\theta_2') + D\cos(\theta_3'))^2 + \cdots \rightarrow} \\
\overline{\leftarrow \cdots (B\sin(\theta_1) + C\sin(\theta_2) + D\sin(\theta_3) + A\cos(\theta_{cross}) + B\cos(\theta_1') + C\cos(\theta_2') + D\cos(\theta_3'))^2}\sin(2\pi f_{RF}t + \theta_4) \\
= -\frac{V_{pk}}{V_\pi L}\pi E(f_{RF})\sin(2\pi f_{RF}t + \theta_4) \\
A = \sqrt{\frac{e^{-\alpha L} + 1 - 2e^{-\frac{\alpha L}{2}}\cos(\frac{2\pi f_{RF}(n_{RF} - n_{o0})L}{c})}{(\frac{2\pi f_{RF}(n_{RF} - n_{o0})}{c})^2 + (\frac{\alpha}{2})^2}} \\
B = \sqrt{\frac{e^{-\alpha L} + 1 - 2e^{-\frac{\alpha L}{2}}\cos(\frac{2\pi f_{RF}(n_{RF} - n_{o1})L}{c})}{(\frac{2\pi f_{RF}(n_{RF} - n_{o1})}{c})^2 + (\frac{\alpha}{2})^2}} \\
C = \sqrt{\frac{e^{-\alpha L} + 1 - 2e^{-\frac{\alpha L}{2}}\cos(\frac{2\pi f_{RF}(n_{RF} - n_{o2})L}{c})}{(\frac{2\pi f_{RF}(n_{RF} - n_{o2})}{c})^2 + (\frac{\alpha}{2})^2}} \\
D = \sqrt{\frac{e^{-\alpha L} + 1 - 2e^{-\frac{\alpha L}{2}}\cos(\frac{2\pi f_{RF}(n_{RF} - n_{o3})L}{c})}{(\frac{2\pi f_{RF}(n_{RF} - n_{o3})}{c})^2 + (\frac{\alpha}{2})^2}} \\
\theta_1 = 2\pi f_{RF}\Delta T_{loop} + \phi_1 - \phi_0, \theta_2 = 4\pi f_{RF}\Delta T_{loop} + \phi_2 - \phi_0, \\
\theta_3 = 6\pi f_{RF}\Delta T_{loop} + \phi_3 - \phi_0, \theta_{cross} = \pi + 6\pi f_{RF}\Delta T_{loop}, \theta_1' = \pi + 8\pi f_{RF}\Delta T_{loop} + \phi_1 - \phi_0, \\
\theta_2' = \pi + 10\pi f_{RF}\Delta T_{loop} + \phi_2 - \phi_0, \theta_3' = \pi + 12\pi f_{RF}\Delta T_{loop} + \phi_3 - \phi_0 \\
\theta_4 = \arctan(\frac{B\sin(\theta_1) + C\sin(\theta_2) + D\sin(\theta_3) + A\cos(\theta_{cross}) + B\cos(\theta_1') + C\cos(\theta_2') + D\cos(\theta_3')}{A + B\cos(\theta_1) + C\cos(\theta_2) + D\cos(\theta_3) + A\cos(\theta_{cross}) + B\cos(\theta_1') + C\cos(\theta_2') + D\cos(\theta_3')}) \\
E(f_{RF}) = \sqrt{(A + B\cos(\theta_1) + C\cos(\theta_2) + D\cos(\theta_3) + A\cos(\theta_{cross}) + B\cos(\theta_1') + C\cos(\theta_2') + D\cos(\theta_3'))^2 + \cdots \rightarrow} \\
\overline{\leftarrow \cdots (B\sin(\theta_1) + C\sin(\theta_2) + D\sin(\theta_3) + A\cos(\theta_{cross}) + B\cos(\theta_1') + C\cos(\theta_2') + D\cos(\theta_3'))^2}
\end{cases}
$$

(15)

By substituting the parameters into Equation(15), the calculated EO bandwidth of the proposed high modulation efficiency phase modulator is shown in **Fig. S12** with a red dotted curve. The measured EO response is described with a blue solid line.

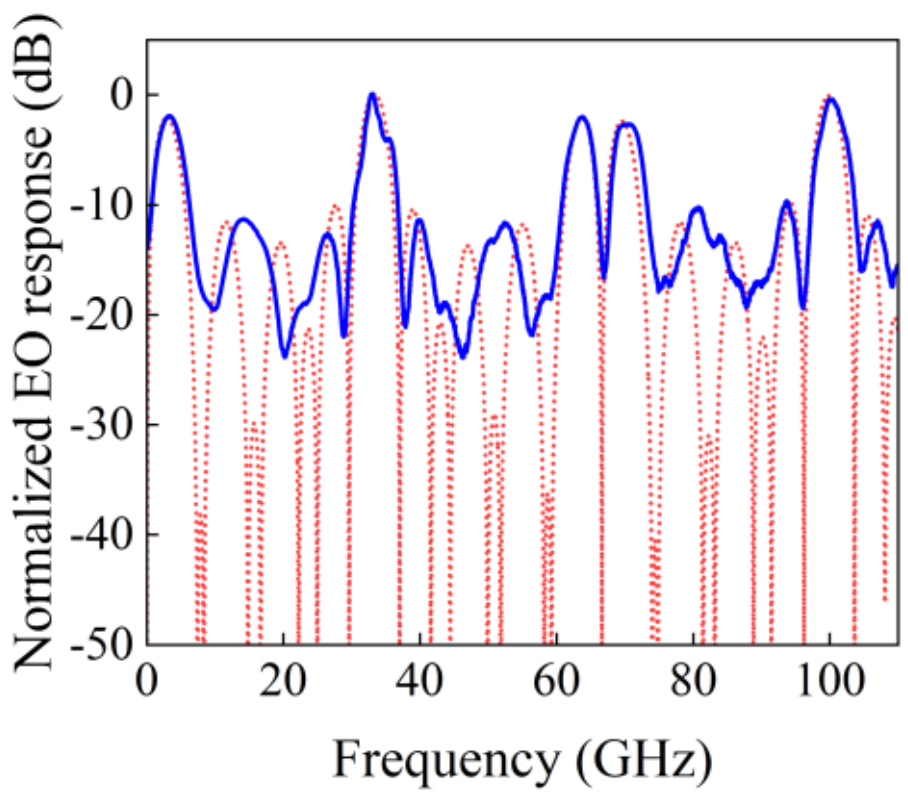

**Figure S12.** Measured (blue solid line) and simulated (red dotted curve) EO bandwidth from 10 MHz to 110 GHz

It can be seen that the simulation results are generally in accordance with the experimental data with the exception of the absence of small sidelobes next to the main resonant peak. We believe that it is influenced by the amplitude and phase responses of spectral fluctuations resulting from the crosstalk between the mode converters and the SMWBs

**Supplementary Note S6: EO response comparison between the NRMs and the common modulators**

From equation (10), we can compare the NRM with three loops and the common modulator without loops. We assume the electrode lengths of them are equal and compare the decay trend of their EO response using our previous setup parameters. The EO response of them is given by the following equations:

$$\begin{cases} \varphi_5(t) = -\frac{V_{pk}}{V_\pi L}\pi\sqrt{(A+B\cos(\theta_1)+C\cos(\theta_2)+D\cos(\theta_3))^2+(B\sin(\theta_1)+C\sin(\theta_2)+D\sin(\theta_3))^2} \\ \qquad \sin(2\pi f_{RF}t+\theta_4) \\ \varphi_6(t) = -\frac{V_{pk}}{V_\pi L}\pi A\sin(2\pi f_{RF}t) \end{cases}$$

(16)

$\varphi_5(t)$ is phase displacement of the NRM with three loops, and $\varphi_6(t)$ is phase displacement of the common modulator without loops. We can define the EO deterioration factor that represent the relative EO deterioration of introducing the additional loops compared to the common modulator without loops. The factor is characterized by:

$$F_{\det} = \left( \frac{-\frac{V_{pk}}{V_\pi L}\pi\sqrt{(A+B\cos(\theta_1)+C\cos(\theta_2)+D\cos(\theta_3))^2+(B\sin(\theta_1)+C\sin(\theta_2)+D\sin(\theta_3))^2}}{-4\frac{V_{pk}}{V_\pi L}\pi A} \right)^2$$
$$= \left( \frac{\sqrt{(1+\frac{B}{A}\cos(\theta_1)+\frac{C}{A}\cos(\theta_2)+\frac{D}{A}\cos(\theta_3))^2+(\frac{B}{A}\sin(\theta_1)+\frac{C}{A}\sin(\theta_2)+\frac{D}{A}\sin(\theta_3))^2}}{4} \right)^2$$

(17)

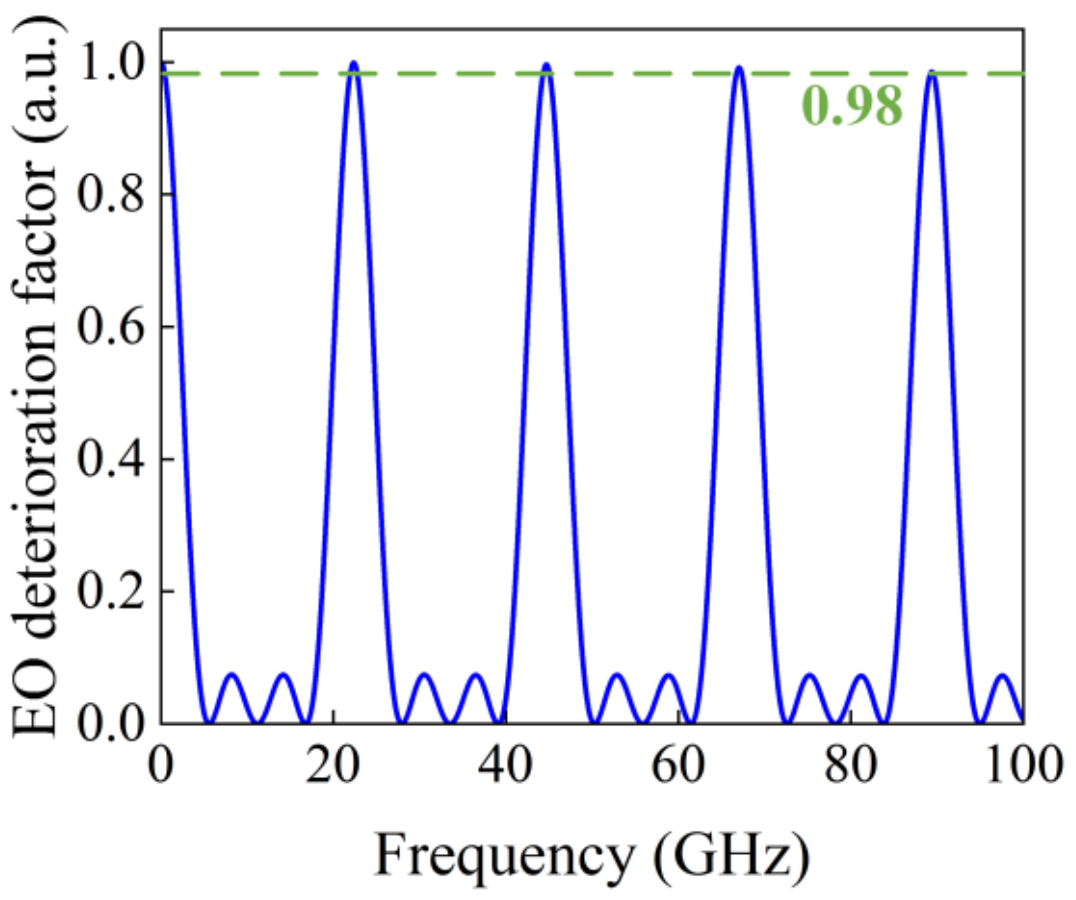

**Figure S13.**Simulated EO deterioration factor from 0 to 100 GHz

In **Figure S13,** we have illustrated a graph depicting the variation of the EO deterioration factor across the frequency range from 0 to 100 GHz. As evident from the graph, all resonant ranges demonstrate values exceeding 0.98, confirming that the EO response of the NRM is virtually indistinguishable from that of a conventional modulator without loops within the resonant frequency range, provided both have the same electrode length. In our research, we have successfully validated that, when the differences in group indices among the utilized modes within the NRM are negligible—specifically, when $A \approx B \approx C \approx D$—both the impact of microwave loss and the index mismatch between optical and microwave signals for the NRM and the common modulator are approximately equivalent.

Although light in the NRM modulator may propagate through multiple electrode lengths, the EO response decay trends remain nearly identical between NRM and conventional modulators with identical electrode lengths. Operating at a currently unattainable high frequency range while maintaining a low $V_\pi$ is not feasible by merely reducing the length of the modulator. Instead, such a goal can be achieved through the implementation of this RM structure.

**Supplementary Note S7: The novel approach to generating microwave photonics rectangular window finite impulse response (FIR) filter.**

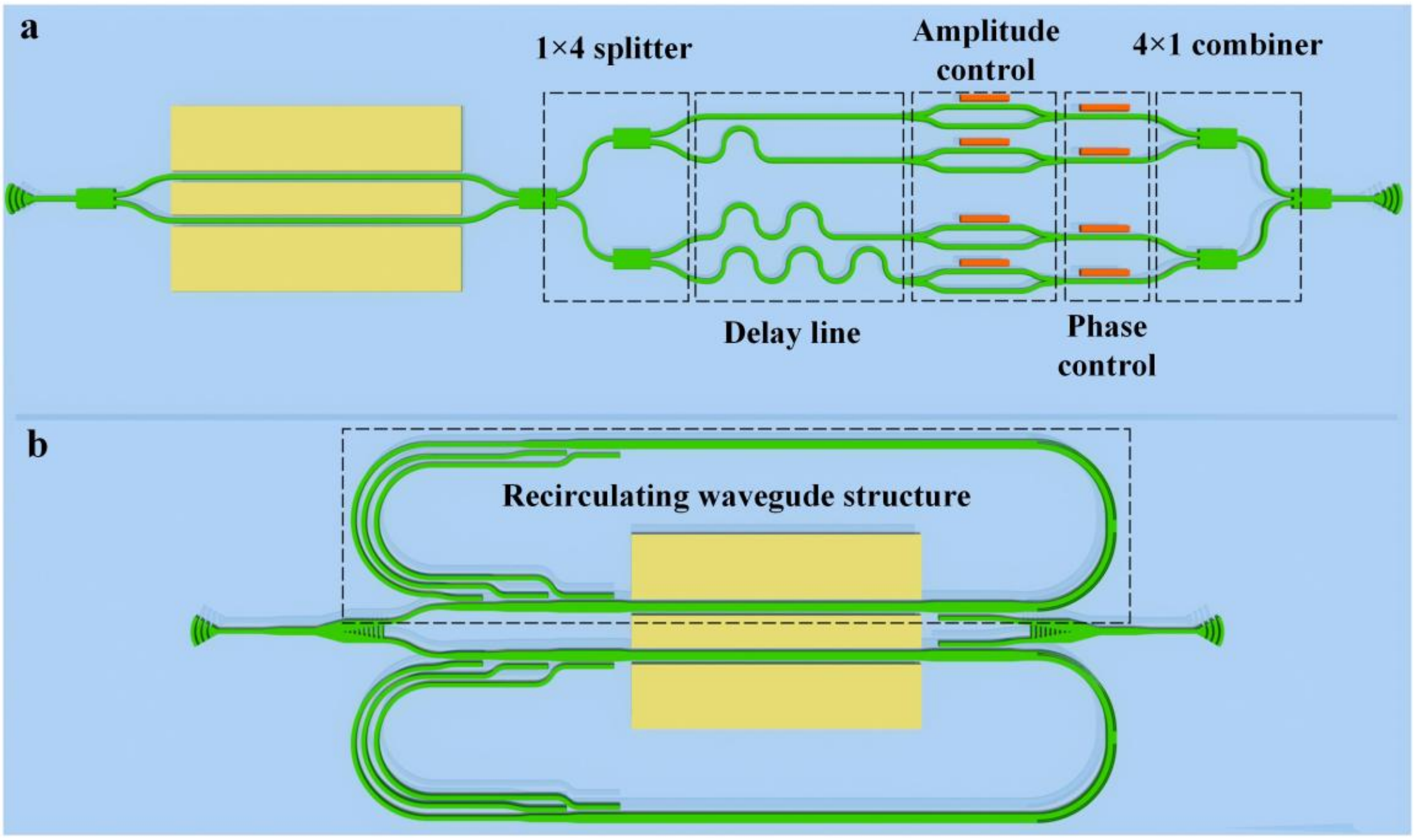

**Figure S14.** The structure comparison of the novel and conventional microwave photonics finite impulse response (FIR) filter **a** the conventional FIR filter, including 1×4 optical splitter, delay lines, amplitude control module, phase control module and 4×1 optical combiner, **b** the novel FIR filter, including novel recirculating waveguide structure.

In conventional N-tap microwave photonic FIR filters, the architecture comprises a MZM operating at its quadrature bias point and N delay lines. Specially, for the 4-tap configuration, when a radio frequency signal $v_m \sin(\omega_m t)$ is applied under small-signal modulation condition, the modulated optical field at the MZM output can be expressed as

$$\begin{aligned} E_{MZM} &= \frac{E_c e^{j\omega_c t}}{2}\left(e^{j\frac{\pi}{2}} e^{jm\sin(\omega_m t)} + e^{-j\frac{\pi}{2}} e^{-jm\sin(\omega_m t)}\right) \\ &= \frac{E_c e^{j\omega_c t}}{2}\left(\sqrt{2}J_0(m) + \sqrt{2}J_1(m)e^{j\omega_m t}e^{j\frac{\pi}{2}} + \sqrt{2}J_1(m)e^{-j\omega_m t}e^{-j\frac{\pi}{2}}\right) \end{aligned} \tag{18}$$

Where $m = \pi \frac{v_m}{v_\pi}$ is the modulation index, $v_m$ is the amplitude of the microwave frequency applied to the MZM, $v_\pi$ is the halfwave voltage of the MZM.

Following the 1×4 optical splitter, each output port undergoes independent phase and amplitude modulation. Prior to recombination in the 4×1 optical combiner, the optical signal in each branch is

$$\begin{cases} E_{port1} = \dfrac{E_c e^{j\omega_c t}}{4} a_1 e^{j\theta_1}(\sqrt{2}J_0(m) + \sqrt{2}J_1(m)e^{j\omega_m t}e^{j\frac{\pi}{2}} + \sqrt{2}J_1(m)e^{-j\omega_m t}e^{-j\frac{\pi}{2}}) \\ E_{port2} = \dfrac{E_c e^{j\omega_c t}e^{j\omega_c \Delta T}}{4} a_2 e^{j\theta_2}(\sqrt{2}J_0(m) + \sqrt{2}J_1(m)e^{j\omega_m t}e^{j\frac{\pi}{2}}e^{j\omega_m \Delta T} + \sqrt{2}J_1(m)e^{-j\omega_m t}e^{-j\frac{\pi}{2}}e^{-j\omega_m \Delta T}) \\ E_{port3} = \dfrac{E_c e^{j\omega_c t}e^{j\omega_c 2\Delta T}}{4} a_3 e^{j\theta_3}(\sqrt{2}J_0(m) + \sqrt{2}J_1(m)e^{j\omega_m t}e^{j\frac{\pi}{2}}e^{j\omega_m 2\Delta T} + \sqrt{2}J_1(m)e^{-j\omega_m t}e^{-j\frac{\pi}{2}}e^{-j\omega_m 2\Delta T}) \\ E_{port4} = \dfrac{E_c e^{j\omega_c t}e^{j\omega_c 3\Delta T}}{4} a_3 e^{j\theta_3}(\sqrt{2}J_0(m) + \sqrt{2}J_1(m)e^{j\omega_m t}e^{j\frac{\pi}{2}}e^{j\omega_m 3\Delta T} + \sqrt{2}J_1(m)e^{-j\omega_m t}e^{-j\frac{\pi}{2}}e^{-j\omega_m 3\Delta T}) \end{cases}$$

(19)

Where $a_1, a_2, a_3, a_4$ are the amplitude coefficient and $\theta_1, \theta_2, \theta_3, \theta_4$ are the phase coefficient of the four branches. For a rectangular window FIR filter, the amplitude coefficients satisfy $a_1 = a_2 = a_3 = a_4 = 1$. To compensate for the optical carrier's phase mismatch induced by the asymmetric branch lengths, the phase coefficients satisfy $\theta_1 = 0, \theta_2 = -\Delta T, \theta_3 = -2\Delta T, \theta_4 = -3\Delta T$. Following signal combination in the 4×1 optical coupler, the resultant optical field is

$$\begin{cases} E_{out} = \dfrac{1}{\sqrt{4}}(E_{port1} + E_{port2} + E_{port3} + E_{port4}) \\ = \dfrac{E_c e^{j\omega_c t}}{8} a_1 e^{j\theta_1}(4\sqrt{2}J_0(m) + \sqrt{2}J_1(m)e^{j\omega_m t}e^{j(\frac{\pi}{2}+\alpha_1)}E_{FIR}(f_{RF}) + \sqrt{2}J_1(m)e^{-j\omega_m t}e^{-j(\frac{\pi}{2}+\alpha_1)}E_{FIR}(f_{RF})) \\ \alpha_1 = \arctan\dfrac{\sin(\omega_m \Delta T) + \sin(2\omega_m \Delta T) + \sin(3\omega_m \Delta T)}{1 + \cos(\omega_m \Delta T) + \cos(2\omega_m \Delta T) + \cos(3\omega_m \Delta T)} \\ E_{FIR}(f_{RF}) = \sqrt{(1 + \cos(\omega_m \Delta T) + \cos(2\omega_m \Delta T) + \cos(3\omega_m \Delta T))^2 + (\sin(\omega_m \Delta T) + \sin(2\omega_m \Delta T) + \sin(3\omega_m \Delta T))^2}) \end{cases}$$

(20)

After PD, the recovered microwave signal with frequency $\omega_m$ is

$$\begin{aligned} I_{ac} &= E_{out} \cdot (E_{out})^* \\ &= \frac{1}{4} R_{PD} E_c{}^2 J_0(m) J_1(m) E_{FIR}(f_{RF}) \cos(\omega_m t + \frac{\pi}{2} + \alpha_1) \end{aligned} \tag{21}$$

where $R_{PD}$ is the PD response. The EO response is then derived by the power of the recovered microwave signal

$$S_{EO}(f_{RF}) = 10 \bullet log_{10} \left| \frac{\frac{1}{4} R_{PD} E_c{}^2 J_0(m) J_1(m) E_{FIR}(f_{RF})}{\frac{1}{4} R_{PD} E_c{}^2 J_0(m) J_1(m) E_{FIR}(0)} \right|^2 = 10 \bullet log_{10} \left| \frac{E_{FIR}(f_{RF})}{E_{FIR}(0)} \right|^2 \tag{22}$$

The EO response of the aforementioned NRM with three loops (11) can be simplified to the following equation under small signal modulation

$$S_{EO}(f_{RF}) == 10 \cdot log_{10}\left|\frac{J_1(-\frac{V_{pk}}{V_\pi L}\pi E(f_{RF}))}{J_1(-\frac{V_{pk}}{V_\pi L}\pi E(0))}\right|^2 = 10 \cdot log_{10}\left|\frac{E(f_{RF})}{E(0)}\right|^2 \tag{23}$$

A comparison between equations (22) and (23) demonstrates that the EO response become identical when $A = B = C = D = 1$ and $\phi_3 - \phi_0 = \phi_2 - \phi_0 = \phi_1 - \phi_0 = 0$ are satisfied

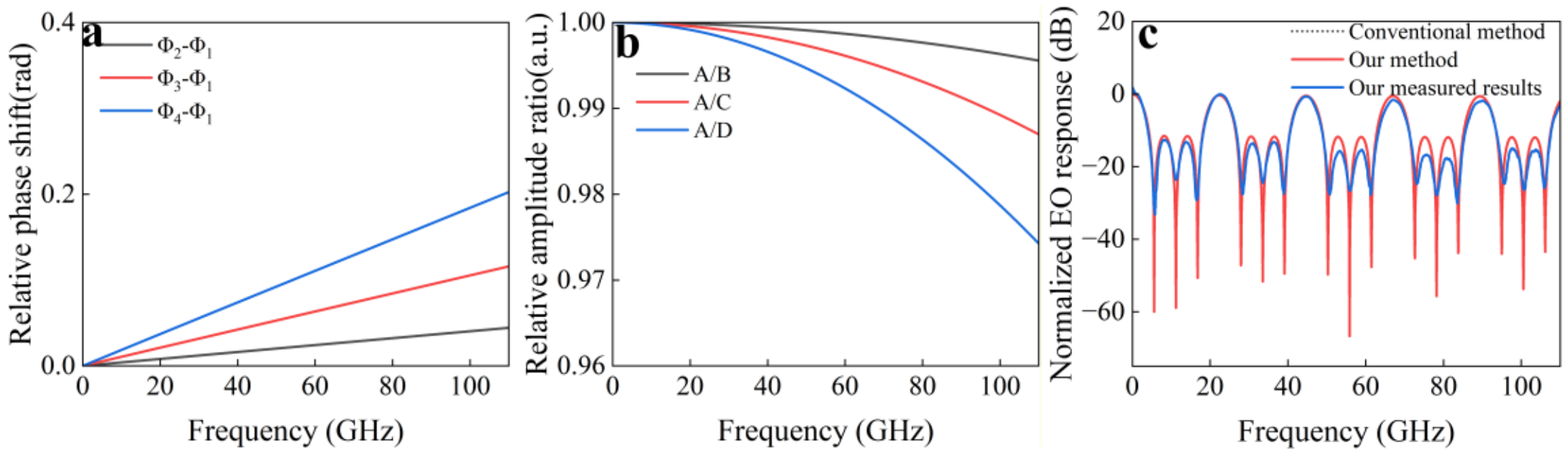

**Figure S15. a** simulated phase offset **b** simulated amplitude ratio offset **c** simulated conventional method (black line) and our method (red line) and our measured results (blue line)

We systematically investigated parameter variations across the 0-110 GHz frequency range. **Figure S15a** demonstrates that the phase offset induced by microwave loss and group index mismatch exhibits a linear frequency-dependent increase, yet remains below 11° throughout the entire spectrum. This minimal phase deviation indicates negligible impact on system performance. **Figure S15b** presents the relative amplitude ratio, which maintains values above 0.97 despite a gradual decline with increasing frequency, further confirming the robustness of the design. Additional performance improvements could potentially be achieved through electrode length optimization or enhanced group index matching.

**Figure S15c** compares our proposed simulation approach (red solid line, equation (11)), the conventional simulation method (black dotted line, simulated by using equation(22)), and experimental measurements (blue solid line). Excellent agreement is observed between our simulation method and conventional calculations. However, the sidelobes exhibit a suppressed passband compared to simulations. This discrepancy may originate from spectral response perturbations caused by crosstalk between the mode converters and the SMWBs.

## Supplementary Table 1. Physical structure parameters of the passive components

| SMWBs | Mode converters | SWG-assisted Y-branch |
|---|---|---|
| $R_0$ = 120 μm | $W_0$ = 1.2 μm $L_1$ = 28 μm | $W_1$ = 3 μm $W_2$ = 0.1 μm |
| $W_4$ = 2.859 μm | $W_1$ = 2.839 μm $L_2$ = 38 μm | $W_3$ = 1 μm $W_4$ = 1 μm |
| $W_5$ = 1.01 μm | $W_2$ = 4.479 μm $L_3$ = 46 μm | $\Lambda$ = 0.3 μm |
| $W_6$ = 0.37 μm $g_1$ = 0.12 μm | $W_3$ = 6.117 μm $g_0$ = 0.2 μm | $d$ = 0.15 μm |

SMWB: slotted multimode waveguide bend. SWG: subwavelength grating.

**Supplementary Table 2: Structures, widths, lengths and group refractive indices for each part in the novel non-resonant recirculating modulator.**

| Structures | Width(μm) | Length(μm) | $TE_0$-$n_g$ | $TE_1$-$n_g$ | $TE_2$-$n_g$ | $TE_3$-$n_g$ |
|---|---|---|---|---|---|---|
| Straight waveguide $L_1$ | 6.117 | 3000 | 2.1906 | 2.2022 | 2.2213 | 2.2477 |
| Straight waveguide $L_2$ | 6.117 | 105 | 2.1906 | 2.2022 | 2.2213 | 2.2477 |
| Straight waveguide $L_3$ | 4.479 | 20 | 2.1935 | 2.2133 | 2.2452 | 2.2838 |
| SMWB $L_4$ | 4.479 | 188.5 | 2.2323 | 2.2511 | 2.2797 | |
| Straight waveguide $L_5$ | 4.479 | 20 | 2.2608 | 2.2815 | 2.3141 | |
| Straight waveguide $L_6$ | 4.479 | 3340.8 | 2.1935 | 2.2133 | 2.2452 | |
| Straight waveguide $L_7$ | 2.839 | 100 | 2.2001 | 2.2403 | | |
| Straight waveguide $L_8$ | 1.2 | 144.8 | 2.2267 | | | |
| Circular waveguide $L_9$ | 1.2 | 396 | 2.261 | | | |
| S bend waveguide $L_{10}$ | 1.2 | 58.8 | 2.2267 | | | |
| Straight waveguide $L_{11}$ | 1.2 | 28 | 2.2267 | | | |
| Straight waveguide $L_{12}$ | 2.839 | 39.155 | 2.2001 | 2.2403 | | |
| Straight waveguide $L_{13}$ | 4.479 | 119.454 | 2.1935 | 2.2133 | 2.2452 | 2.2838 |
| Straight waveguide $L_{14}$ | 6.117 | 155.936 | 2.1906 | 2.2022 | 2.2213 | 2.2477 |
| S bend waveguide $L_{15}$ | 1.2 | 76.77 | 2.2267 | | | |
| Straight waveguide $L_{16}$ | 1.2 | 119.295 | 2.2267 | | | |
| Circular waveguide $L_{17}$ | 1.2 | 366.21 | 2.261 | | | |
| Straight waveguide $L_{18}$ | 1.2 | 102.1 | 2.2267 | | | |
| S bend waveguide $L_{19}$ | 1.2 | 84.37 | 2.2267 | | | |
| Straight waveguide $L_{20}$ | 1.2 | 38 | 2.2267 | | | |
| S bend waveguide $L_{21}$ | 1.2 | 126.3 | 2.2267 | | | |
| Straight waveguide $L_{22}$ | 1.2 | 194.628 | 2.2267 | | | |
| Circular waveguide $L_{23}$ | 1.2 | 309.155 | 2.261 | | | |
| Straight waveguide $L_{24}$ | 1.2 | 186.186 | 2.2267 | | | |
| S bend waveguide $L_{25}$ | 1.2 | 135.54 | 2.2267 | | | |
| Straight waveguide $L_{26}$ | 1.2 | 46 | 2.2267 | | | |
| Waveguide width $W_1$ | 6.117 | | 2.1906 | 2.2022 | 2.2213 | 2.2477 |
| Waveguide width $W_2$ | 4.479 | | 2.1906 | 2.2022 | 2.2213 | |
| Waveguide width $W_3$ | 2.839 | | 2.1906 | 2.2022 | | |
| Waveguide width $W_4$ | 1.2 | | 2.1906 | | | |
| Taper $T_1$ | 6.117-4.479 | 30 | 2.1921 | 2.2078 | 2.2333 | |
| Taper $T_2$ | 4.479-2.839 | 30 | 2.1968 | 2.2268 | | |
| Taper $T_3$ | 2.839-1.2 | 30 | 2.2134 | | | |

SMWB: slotted multimode waveguide bend.

**Supplementary Table 3. Structures, widths, lengths and group refractive indices for each part in the Mach-Zehnder modulator.**

| Structures | Width(μm) | Length(μm) | $TE_0$-ng | $TE_1$-ng | $TE_2$-ng | $TE_3$-ng |
|---|---|---|---|---|---|---|
| Straight waveguide $L_1$ | 4.479 | 2000 | 2.1935 | 2.2133 | 2.2452 | 2.2838 |
| Straight waveguide $L_2$ | 6.117 | 105 | 2.1906 | 2.2022 | 2.2213 | 2.2477 |
| Straight waveguide $L_3$ | 4.479 | 20 | 2.1935 | 2.2133 | 2.2452 | 2.2838 |
| SMWB $L_4$ | 4.479 | 188.5 | 2.2323 | 2.2511 | 2.2797 | |
| Straight waveguide $L_5$ | 4.479 | 20 | 2.2608 | 2.2815 | 2.3141 | |
| Straight waveguide $L_6$ | 4.479 | 2340.8 | 2.1935 | 2.2133 | 2.2452 | |
| Straight waveguide $L_7$ | 2.839 | 100 | 2.2001 | 2.2403 | | |
| Straight waveguide $L_8$ | 1.2 | 144.8 | 2.2267 | | | |
| Circular waveguide $L_9$ | 1.2 | 396 | 2.2267 | | | |
| S bend waveguide $L_{10}$ | 1.2 | 58.8 | 2.2267 | | | |
| Straight waveguide $L_{11}$ | 1.2 | 28 | 2.2267 | | | |
| Straight waveguide $L_{12}$ | 2.839 | 39.155 | 2.2001 | 2.2403 | | |
| Straight waveguide $L_{13}$ | 4.479 | 119.454 | 2.1935 | 2.2133 | 2.2452 | 2.2838 |
| Straight waveguide $L_{14}$ | 6.117 | 125.936 | 2.1906 | 2.2022 | 2.2213 | 2.2477 |
| S bend waveguide $L_{15}$ | 1.2 | 76.77 | 2.2267 | | | |
| Straight waveguide $L_{16}$ | 1.2 | 119.02 | 2.2267 | | | |
| Circular waveguide $L_{17}$ | 1.2 | 366.21 | 2.261 | | | |
| Straight waveguide $L_{18}$ | 1.2 | 101.53 | 2.2267 | | | |
| S bend waveguide $L_{19}$ | 1.2 | 84.37 | 2.2267 | | | |
| Straight waveguide $L_{20}$ | 1.2 | 38 | 2.2267 | | | |
| S bend waveguide $L_{21}$ | 1.2 | 126.3 | 2.2267 | | | |
| Straight waveguide $L_{22}$ | 1.2 | 198.94 | 2.2267 | | | |
| Circular waveguide $L_{23}$ | 1.2 | 309.155 | 2.261 | | | |
| Straight waveguide $L_{24}$ | 1.2 | 190.58 | 2.2267 | | | |
| S bend waveguide $L_{25}$ | 1.2 | 135.54 | 2.2267 | | | |
| Straight waveguide $L_{26}$ | 1.2 | 46 | 2.2267 | | | |
| Waveguide width $W_1$ | 4.479 | | 2.1906 | 2.2022 | 2.2213 | |
| Waveguide width $W_2$ | 6.117 | | 2.1906 | 2.2022 | 2.2213 | 2.2477 |
| Waveguide width $W_3$ | 2.839 | | 2.1906 | 2.2022 | | |
| Waveguide width $W_4$ | 1.2 | | 2.1906 | | | |
| Taper $T_1$ | 6.117-4.479 | 30 | 2.1921 | 2.2078 | 2.2333 | |
| Taper $T_2$ | 4.479-2.839 | 30 | 2.1968 | 2.2268 | | |
| Taper $T_3$ | 2.839-1.2 | 30 | 2.2134 | | | |

SMWB: slotted multimode waveguide bend.

**Supplementary Table 4. Structures, widths, lengths and group refractive indices for each part in the high modulation efficiency phase modulator type II**

| Structures | Width(µm) | Length(µm) | $TE_0$-ng | $TE_1$-ng | $TE_2$-ng | $TE_3$-ng |
|---|---|---|---|---|---|---|
| Straight waveguide $L_1$ | 4.479 | 1000 | 2.1935 | 2.2133 | 2.2452 | 2.2838 |
| Straight waveguide $L_2$ | 6.117 | 105 | 2.1906 | 2.2022 | 2.2213 | 2.2477 |
| Straight waveguide $L_3$ | 4.479 | 20 | 2.1935 | 2.2133 | 2.2452 | 2.2838 |
| SMWB $L_4$ | 4.479 | 188.5 | 2.2323 | 2.2511 | 2.2797 | |
| Straight waveguide $L_5$ | 4.479 | 20 | 2.2608 | 2.2815 | 2.3141 | |
| Straight waveguide $L_6$ | 4.479 | 1340.8 | 2.1935 | 2.2133 | 2.2452 | |
| Straight waveguide $L_7$ | 2.839 | 100 | 2.2001 | 2.2403 | | |
| Straight waveguide $L_8$ | 1.2 | 144.8 | 2.2267 | | | |
| Circular waveguide $L_9$ | 1.2 | 396 | 2.261 | | | |
| S bend waveguide $L_{10}$ | 1.2 | 58.8 | 2.2267 | | | |
| Straight waveguide $L_{11}$ | 1.2 | 28 | 2.2267 | | | |
| Straight waveguide $L_{12}$ | 2.839 | 39.155 | 2.2001 | 2.2403 | | |
| Straight waveguide $L_{13}$ | 4.479 | 119.454 | 2.1935 | 2.2133 | 2.2452 | 2.2838 |
| Straight waveguide $L_{14}$ | 6.117 | 125.936 | 2.1906 | 2.2022 | 2.2213 | 2.2477 |
| S bend waveguide $L_{15}$ | 1.2 | 76.77 | 2.2267 | | | |
| Straight waveguide $L_{16}$ | 1.2 | 128.19 | 2.2267 | | | |
| Circular waveguide $L_{17}$ | 1.2 | 366.21 | 2.261 | | | |
| Straight waveguide $L_{18}$ | 1.2 | 111 | 2.2267 | | | |
| S bend waveguide $L_{19}$ | 1.2 | 84.37 | 2.2267 | | | |
| Straight waveguide $L_{20}$ | 1.2 | 38 | 2.2267 | | | |
| S bend waveguide $L_{21}$ | 1.2 | 126.3 | 2.2267 | | | |
| Straight waveguide $L_{22}$ | 1.2 | 222.36 | 2.2267 | | | |
| Circular waveguide $L_{23}$ | 1.2 | 309.155 | 2.261 | | | |
| Straight waveguide $L_{24}$ | 1.2 | 213.9 | 2.2267 | | | |
| S bend waveguide $L_{25}$ | 1.2 | 135.54 | 2.2267 | | | |
| Straight waveguide $L_{26}$ | 1.2 | 46 | 2.2267 | | | |
| Straight waveguide $L_{27}$ | 1.2 | 49.3 | 2.2267 | | | |
| Circular waveguide $L_{28}$ | 1.2 | 314.159 | 2.261 | | | |
| Straight waveguide $L_{29}$ | 1.2 | 183.283 | 2.2267 | | | |
| Straight waveguide $L_{30}$ | 1.2 | 122.41 | 2.2953 | | | |
| Straight waveguide $L_{31}$ | 1.2 | 1963.25 | 2.2267 | | | |
| Straight waveguide $L_{32}$ | 1.2 | 74.01 | 2.2953 | | | |
| Straight waveguide $L_{33}$ | 1.2 | 150 | 2.2267 | | | |
| Waveguide width $W_1$ | 4.479 | | 2.1906 | 2.2022 | 2.2213 | |
| Waveguide width $W_2$ | 6.117 | | 2.1906 | 2.2022 | 2.2213 | 2.2477 |
| Waveguide width $W_3$ | 2.839 | | 2.1906 | 2.2022 | | |
| Waveguide width $W_4$ | 1.2 | | 2.1906 | | | |
| Taper $T_1$ | 6.117-4.479 | 30 | 2.1921 | 2.2078 | 2.2333 | |
| Taper $T_2$ | 4.479-2.839 | 30 | 2.1968 | 2.2268 | | |
| Taper $T_3$ | 2.839-1.2 | 30 | 2.2134 | | | |
| Taper $T_4$ | 2.839-1.2 | 50 | 2.2134 | | | |

SMWB: slotted multimode waveguide bend.

**Supplementary References:**